\documentclass[12pt]{article}

\usepackage{amsfonts}
\usepackage{amsmath}
\usepackage{amssymb}
\usepackage{amsthm}
\usepackage{youngtab}
\usepackage{braket}
\usepackage{graphicx}%\usepackage{cite}	%or \usepackage[dvips]{graphicx}%\usepackage{cite}
\usepackage[svgnames]{xcolor}			%or \usepackage[dvips,x11names]{xcolor}
\usepackage{ytableau,tikz}					%or \usepackage{tikz}
\usepackage{here}
\usepackage{comment}
\usepackage{color}                    % For creating coloured text and background
\usepackage[colorlinks=true,linkcolor=black,citecolor=black,urlcolor=black]{hyperref}    % For creating hyperlinks in cross references    
\usetikzlibrary{intersections, calc,shapes.geometric,patterns,arrows,decorations.pathreplacing} 	% library for tikz 
\newcommand{\ba}{\begin{eqnarray}}
\newcommand{\ea}{\end{eqnarray}}
\newcommand{\nn}{\nonumber}
\newcommand{\rrangle}{\rangle}

\newcommand{\tn}{{\tilde n}}
\newcommand{\tm}{{\tilde m}}

\newcommand{\cecm}{{\check m}}
\newcommand{\cecn}{{\check n}}
\newcommand{\ceci}{{\check i}}
\newcommand{\cecj}{{\check j}}
\newcommand{\ceck}{{\check k}}
\newcommand{\cC}{\mathcal{C}}
\newcommand{\cD}{\mathcal{D}}
\newcommand{\cW}{\mathcal{W}}

\makeatletter
\@addtoreset{equation}{section}

\makeatother

\def\thefootnote{\ifnum\c@footnote>\z@\textasteriskcentered\@arabic\c@footnote\fi}
\makeatletter
\renewcommand{\footnoterule}{%
\kern-3\p@
\hrule width 0.4\columnwidth
\kern 2.6\p@}
\def\thefootnote{\ifnum\c@footnote>\z@\@arabic\c@footnote\fi}
\makeatother
\makeatletter
\newcommand{\@authornote}[2]{{\def\thefootnote{\fnsymbol{footnote}}\setcounter{footnote}{#1}#2\setcounter{footnote}{0}}}
\newcommand{\authornotemark}[1]{\@authornote#1{\addtocounter{footnote}{-1}\footnotemark}}
\newcommand{\authornotetext}[2]{\@authornote#1{\footnotetext{#2}}}
\makeatother

\begin{document}
\theoremstyle{theorem}
\newtheorem*{theorem}{Theorem}
\newtheorem*{conjecture}{Conjecture}
\newtheorem*{proposition}{Proposition}
\newtheorem*{property}{Property}
\newtheorem*{corollary}{Corollary}
\newtheorem*{lemma}{Lemma}
\newtheorem*{claim}{Claim}
\newtheorem*{observation}{Observation}
\theoremstyle{definition}
\newtheorem*{definition}{Definition}
\newtheorem*{example}{Example}
\theoremstyle{remark}
\newtheorem*{remark}{Remark}
\newcommand{\bsq}{\hspace{\fill}$\blacksquare$}
\begin{titlepage}

\begin{flushright}
UT-15-32
\end{flushright}

\vskip 12mm

\begin{center}
{\Large\bf SH$^c$ Realization of Minimal Model CFT:\\
 Triality, Poset and Burge Condition}
\vskip 2cm
{\Large M. Fukuda, S. Nakamura, Y. Matsuo and R.-D. Zhu}
\vskip 2cm
{\it Department of Physics, The University of Tokyo}\\
{\it Hongo 7-3-1, Bunkyo-ku, Tokyo 113-0033, Japan}
\end{center}
\vfill
\begin{abstract}
Recently an orthogonal basis of $\cW_N$-algebra (AFLT basis) labeled by $N$-tuple 
Young diagrams was found in the context of 4D/2D duality.
Recursion relations among the basis are summarized in the form of
an algebra SH$^c$ which is universal for any $N$.
%It includes an infinite number of commuting operators which
%are diagonal on the basis and 
We show that it has an $\mathfrak{S}_3$ automorphism
which is referred to as triality.
We study the  level-rank duality between minimal models, %from SH$^c$
which is a special example of the automorphism. %which holds among minimal model CFTs.
It is shown that the nonvanishing states in both systems are described by $N$ or
$M$ Young diagrams with the rows of boxes appropriately shuffled.
%The analysis demonstrates that SH$^c$ has triality symmetry
%for some specific choices of parameters.  
The reshuffling of rows implies there exists partial ordering
of the set which labels them.  
For the simplest example, one can compute the partition functions for the partially ordered set (poset)
explicitly, which reproduces the Rogers-Ramanujan identities.
We also study the description of minimal models by SH$^c$.
Simple analysis reproduces some known properties of minimal models,
the structure of singular vectors and the $N$-Burge condition in the Hilbert space. 
\end{abstract}
\vfill
\end{titlepage}

\renewcommand{\thefootnote}{\dag\arabic{footnote}}
\setcounter{footnote}{0}
%%%%%
\ytableausetup{centertableaux,boxsize=.6em}
%%%%%
%\noindent\hrulefill
%\tableofcontents
%\noindent\hrulefill
\section{Introduction}
%%%%%%%%
Many years ago \cite{Fateev:1987vh}, 
$\cW$-algebra was formulated as a nonlinear generalization of
the two dimensional conformal field theory and has been playing
significant roles in many branches of physics, such as
string theory, quantum gravity, the statistical mechanics,
and the exactly solvable systems.
It was defined as an extended conformal symmetry
with higher spin currents.  In the case of $\cW_N$-algebra, the generators
of the symmetry consist of spin $2, 3, \cdots, N$ currents.
The commutation relations among them are in general nonlinear
and the explicit form of the algebra is known only for simple cases such as $\cW_3$.
For the general cases, while the algebra itself is not given explicitly, 
the representation was derived through
the realization with free bosons \cite{Fateev:1987zh}.

Recently, in the efforts to prove the 4D/2D correspondence \cite{Alday2010}
a new realization of A-type $\cW$-algebra was constructed
\cite{schiffmann2013cherednik,Kanno:2013aha}.
It is based on the orthogonal AFLT basis 
(Alba-Fateev-Litvinov-Tarnopolskiy basis) labeled by an array of Young diagrams
\cite{Alba2011,Fateev:2011hq}.
The algebra has a rather lengthy name (central extension of
spherical degenerate double affine
Hecke algebra) but was abbreviated to SH$^c$ in \cite{schiffmann2013cherednik},
which will be used in this paper.
The generators of the algebra, $D_{r,l}$, are labeled by two integers\footnote{We use the convention $0\in\mathbb{N}$.}, 
$r\in \mathbb{Z}$ and $l\in \mathbb{N}$.
The essential part of the algebra is written explicitly in terms of
$D_{\pm 1, l}$ and
$D_{0,l}$.

The rank $N$ representation of SH$^c$ is spanned by $N$-tuple Young diagrams
and can be realized
in terms of $N$ free bosons. 
The explicit forms of the generators are
given for $D_{r,0}$ and $D_{r,1}$.
The generators $D_{r,0}$
are identified with free $U(1)$ current
 and some combinations of $D_{r,0}$ and $D_{r,1}$ are identified
as the Virasoro generators in $\cW_N$-algebra.
The other $\cW_N$ currents are given in terms of higher generators
(in less manifest fashion) in \cite{schiffmann2013cherednik, Matsuo:2014rba}.
In this way, the generators of two algebras can be identified through
the bosonic oscillators.  We note that
$N$ is arbitrary while the algebra is the same.
In this sense, SH$^c$ may be regarded
as a universal $\cW$-symmetry which contains representations of
all  $\cW_N$-algebra.

The purpose of the paper is to find more direct link
between the representations of $\cW_N$-algebra and SH$^c$.
Our first focus is the explicit realization of the level-rank duality
\cite{kuniba1991ferro, Altschuler:1990th}  in the SH$^c$ module.
It is a duality between minimal models in $\cW_N$- and $\cW_M$-algebras
with $N\neq M$.
This is somehow puzzling since we need to find a direct correspondence
between two Hilbert spaces spanned by different number of Young diagrams.
Another motivation to study the duality is that  it would be related to the so-called
triality symmetry of $\cW_{\infty}[\mu]$ \cite{Gaberdiel:2011wb,Gaberdiel:2012ku}.
In our notation, it is realized by a discrete non-abelian automorphism
$\mathfrak{S}_3$ through the  transformation of a parameter $\beta$.
It is generated by an obvious transformation $\beta\leftrightarrow 1/\beta$
and less obvious one $\beta\leftrightarrow \frac{\beta}{\beta-1}$.
%The second is the level-rank duality when $\beta$ is taken to a specific form.
We show that this automorphism holds in SH$^c$ in general.
The level-rank duality is interesting since they are realized among the 
minimal model CFTs where we need to impose $\beta$ to take a specific form.

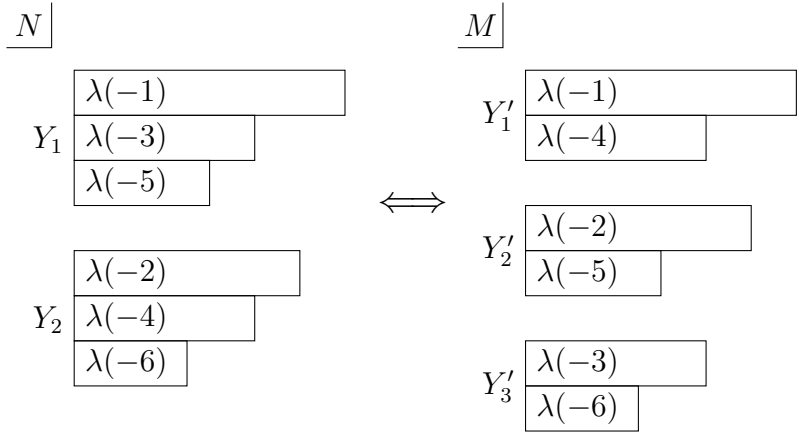
\begin{figure}
\centering
\begin{tikzpicture}[scale=.6]
\coordinate (A)at(0,0);
\node[right] at([yshift=-.5cm]A){$\lambda(-1)$};
\draw (A) rectangle ([xshift=6cm,yshift=-1cm]A);
\coordinate (B)at(0,-1);
\node[right] at([yshift=-.5cm]B){$\lambda(-3)$};
\draw (B) rectangle ([xshift=4cm,yshift=-1cm]B);
\coordinate (C)at(0,-2);
\node[right] at([yshift=-.5cm]C){$\lambda(-5)$};
\draw (C) rectangle ([xshift=3cm,yshift=-1cm]C);
\coordinate (D)at(0,-4);
\node[right] at([yshift=-.5cm]D){$\lambda(-2)$};
\draw (D) rectangle ([xshift=5cm,yshift=-1cm]D);
\coordinate (E)at(0,-5);
\node[right] at([yshift=-.5cm]E){$\lambda(-4)$};
\draw (E) rectangle ([xshift=4cm,yshift=-1cm]E);
\coordinate (F)at(0,-6);
\node[right] at([yshift=-.5cm]F){$\lambda(-6)$};
\draw (F) rectangle ([xshift=2.5cm,yshift=-1cm]F);
\node[] at (-1,1) {$N$};
\node[] at (3,1) {};
\node[below left] at (B) {$Y_1$};
\node[below left] at (E) {$Y_2$};
%\draw[fill=gray!50] (0,-4) rectangle (10,-5);
%\draw[fill=gray!70] (0,-4) rectangle (1,-5);
\draw (-1.5,.5)--(-.5,.5)--(-.5,1.5);
%\draw[->,bend right,ultra thick, in=-90, out=-90] ([yshift=2.5cm]1,-1)node[right]{A row labeled by $x\in\mathbb{Z}$}to([xshift=.5cm,yshift=-.5cm]0,-4);
\node[] at (7.5, -3) {\large$\Longleftrightarrow$};
\coordinate (A1)at(10,0);
\node[right] at([yshift=-.5cm]A1){$\lambda(-1)$};
\draw (A1) rectangle ([xshift=6cm,yshift=-1cm]A1);
\coordinate (B1)at(10,-3);
\node[right] at([yshift=-.5cm]B1){$\lambda(-2)$};
\draw (B1) rectangle ([xshift=5cm,yshift=-1cm]B1);
\coordinate (C1)at(10,-6);
\node[right] at([yshift=-.5cm]C1){$\lambda(-3)$};
\draw (C1) rectangle ([xshift=4cm,yshift=-1cm]C1);
\coordinate (D1)at(10,-1);
\node[right] at([yshift=-.5cm]D1){$\lambda(-4)$};
\draw (D1) rectangle ([xshift=4cm,yshift=-1cm]D1);
\coordinate (E1)at(10,-4);
\node[right] at([yshift=-.5cm]E1){$\lambda(-5)$};
\draw (E1) rectangle ([xshift=3cm,yshift=-1cm]E1);
\coordinate (F1)at(10,-7);
\node[right] at([yshift=-.5cm]F1){$\lambda(-6)$};
\draw (F1) rectangle ([xshift=2.5cm,yshift=-1cm]F1);
\node[] at (9,1) {$M$};
\draw (8.5,.5)--(9.5,.5)--(9.5,1.5);
\node[] at (13,1) {};
\node[left] at (D1) {$Y'_1$};
\node[left] at (E1) {$Y'_2$};
\node[left] at (F1) {$Y'_3$};
\end{tikzpicture}
\caption{An example of reshuffling rows as a realization of the level-rank duality
($N=2, M=3$, %$\Delta=-\frac15$
). We take the set of labels $X=\{-1,-2,\cdots\}$ which represents the rows.
We put $\lambda(-7)=\lambda(-8)=\cdots=0$ in this example.
}\label{shuffle:intro}
\end{figure}

We note that the AFLT basis is diagonal with respect to infinite commuting operators
in SH$^c$.  The eigenvalues are the power sum of  numbers assigned
to each box of the Young diagrams.  By analyzing the integer assigned to each
box, we find that the two Hilbert spaces are related by reshuffling rows
contained in each Young diagrams (see Figure \ref{shuffle:intro}).
To be specific, we show that there is a set of
numbers $X$ which labels the rows of Young diagrams.  The set $X$ is determined
by the representation of $\cW_N$ ($\cW_M$).
For each $x\in X$, we assign a positive number $\lambda(x)$
which satisfies both $\lambda(x)\geq \lambda(x-N)$
and $\lambda(x)\geq\lambda(x-M)$.
We show that the $N$-(resp. $M$-) tuple Young diagrams $\vec Y$  (resp. $\vec Y'$)
in $\cW_N$ (resp. $\cW_M$) is described by such  a partition.

The counting of states with such property was studied in the literature \cite{Stanley:1986:EC:21786}
which is referred as the partition function of poset (partially ordered set).
In the simplest case $(N,M)=(2,3)$, one can solve the combinatorics completely to obtain
the formula for the partition function.  It turns out that the summation takes the
same form as Rogers-Ramanujan formulae and agree with the known character formula
(for example \cite{Altschuler:1990th}) of the minimal models of the $\cW$-algebra.

The Hilbert space for SH$^c$ should coincide with that
of $\cW_N$ module.  To confirm it, we
also study the description of the Hilbert space for general minimal models.
In particular, we show that the singular vectors in the Hilbert space
have a simple graphical interpretation and they
agree with those for $\cW_N$ module.
We also demonstrate explicitly that the condition which
characterizes the Hilbert space of the minimal models,
the $N$-Burge condition which was proposed recently \cite{Belavin:2015ria},
by showing that any states which violate the condition are not produced
by the operators in SH$^c$.

This paper is organized as follows. In section \ref{section2}, we review some 
basic aspects of minimal models of $\cW_N$-algebra, the level-rank duality and
the triality.  We explain the correspondence of the conformal dimensions in
some detail to prepare the notation in the following sections.
A brief summary of the SH$^c$ is given in section \ref{section3}. 
We show how the triality is realized in SH$^c$.  It is an exact automorphism of
the algebra.
%as long as the central charges are trivial. If they are nonvanishing, the second
%transformation does not preserve the algebra in general.  
In section \ref{section4}, we describe the Hilbert space of the minimal models
by SH$^c$.   We show that the null states appearing in the Hilbert space agree
with those for $\cW_N$ module.
 In section \ref{section5}, we prove that the central charges of SH$^c$
 agree for a pair of minimal models with the level-rank duality.
%It implies that the triality symmetry is realized for these nontrivial cases.
While the triality is an exact automorphism of SH$^c$ in general, 
this example is special since it is realized through the tuning of finite parameters.
In section \ref{section6}, we demonstrate the duality at the state-to-state level. 
The correspondence between the partition of the poset 
and Young diagrams mentioned above is explained in detail.
In particular we prove the partition for poset gives the Hilbert space correctly.
In section \ref{section6a}, we explain how to calculate 
the partition function for the poset according to \cite{Stanley:1986:EC:21786}.
For the simplest case $(2,3)$ the computation can be performed exactly and
equality with the know partition function gives the Rogers-Ramanuman identity.
In section \ref{section7}, we come back to the general minimal models.
We study the action of generators of SH$^c$ and derive the $N$-Burge condition.

%%%%%%%%%%%%%%%%%%%%%%%%%%
\section{Brief review of level-rank duality in $\cW$-algebras}\label{section2}
In this section, we recall some basic facts about minimal models and the level-rank duality that will be important later. 
Especially a way to identify the dual primary fields for
 the level-rank duality pair is given at the end of this section in detail and we will see  
 a similar duality in SH$^c$ in later sections.
\subsection{Minimal models of $\cW_N$-algebra}
A minimal model of the $\cW_N$-algebra \cite{Fateev:1987zh,Bouwknegt:1992wg} is characterized by a central charge,
\ba
c=(N-1)\left(
1-\frac{(p-q)^2}{pq} N(N+1)
\right)\,,
\ea
which is parametrized by a pair of coprime positive integers $(p,q)$.
The highest weights of primary fields are limited to
\ba\label{weight}
\Delta(n_1,\dots, n_{N-1}; n'_1,\dots,n'_{N-1})
=\frac{12\left( \sum_{i=1}^{N-1}(p n_i- q n'_i)\vec\omega_i\right)^2-N(N^2-1) (p-q)^2}{24pq}\,,
\ea
where two sequences $(n_i)_{i=1}^{N-1}$ and $(n'_i)_{i=1}^{N-1}$ consist of positive integers subjected to the following inequalities,
\ba\label{cnst}
\sum_{i=1}^{N-1}  n_i\leq q-1, \quad \sum_{i=1}^{N-1} n'_i\leq p-1,
\ea
and $\{\vec \omega_i\}_{i=1}^{N-1}$ are the fundamental weights of $\mathfrak{su}(N)$ satisfying,
\ba\label{omegaproduct}
\vec \omega_i\cdot \vec \omega_j=\frac{i(N-j)}{N} ,\quad i\leq j\,.
\ea
The minimal model is realized by the coset construction in terms of $\mathrm{SU}(N)$ current algebra,
\begin{equation}
\mathcal{W}_{N,k}\equiv \frac{\mathfrak{su}(N)_k\oplus \mathfrak{su}(N)_1}{\mathfrak{su}(N)_{k+1}}\,,
\end{equation}
with the level $k$ given as
\begin{equation}\label{kpq}
k=\frac{p}{q-p}-N\,.
\end{equation}
This model is unitary when $k$ is a non-negative integer. 

The Hilbert space associated with a highest weight $\Delta((n_{i}), (n'_{i}))$ is 
completely degenerate in a sense that there are $N$ singular vectors appearing at the levels 
$n_i n'_i$ ($1\le i\le N$), where $n_N\equiv q-\sum_{i=1}^{N-1} n_i$ and 
$n'_N\equiv p-\sum_{i=1}^{N-1} n'_i$. 
Note that the inequalities (\ref{cnst}) ensure that $n_{N}$ and $n'_{N}$ are positive integers and we have
\begin{gather}
\sum_{i=1}^{N}n_{i}=q,\  \sum_{i=1}^{N}n'_{i}=p. 
\end{gather}

%%%%%%%%%%%%%%%%%%%%%%%%%
\subsection{Level-Rank duality}\label{LRdual}
%%%%%%%%%%%%%%%%%%%%%%%%%
The level-rank duality is an extra symmetry
which interpolates a $\cW_N$-algebra with another $\cW_{M}$-algebra
\cite{kuniba1991ferro, Altschuler:1990th} when the parameters of the minimal model
$p,q$ take special values.
It is written as a correspondence between coset models:
\begin{equation}
\mathcal{W}_{N,k}=\frac{\mathfrak{su}(N)_k\oplus \mathfrak{su}(N)_1}{\mathfrak{su}(N)_{k+1}}
\sim \frac{\mathfrak{su}(M)_l\oplus \mathfrak{su}(M)_1}{\mathfrak{su}(M)_{l+1}}=\mathcal{W}_{M,l}\,,
\end{equation}
where $N$ and $M$ are coprime and the levels $k$, $l$ are given as
\begin{equation}
k=\frac{N}{M}-N, \quad l=\frac{M}{N}-M\,.
\end{equation}
Both models have the same central charge:
\begin{align}
(N-1)\left(1-\frac{N(N+1)}{(N+k)(N+k+1)}\right)
&=(M-1)\left(1-\frac{M(M+1)}{(M+l)(M+l+1)}\right)\nonumber\\
&=
-\frac{(N-1)(M-1)(N+M+NM)}{N+M}\,,
\end{align}
We can see from (\ref{kpq}) that the model $\mathcal{W}_{N,k}$ corresponds to the one with $p=N$ and $q=N+M$.
Since $k\notin\mathbb{N}$, the model is not unitary.

The inequalities (\ref{cnst}) imply that $n'_i=1$ for all $1\le i\le N-1$. 
We introduce the notation $\tn_i\equiv n_i-1\geq 0$, which are then constrained by $\sum_{i=1}^{N-1}\tn_i\leq M$, 
and denote highest weights by
\ba
\Delta(\tn)\equiv\Delta(n_1,\cdots, n_{N-1}; 1,\cdots,1)\,.
\ea
Note that the conformal dimension $\Delta(\tn)$ is simplified to
\ba
\Delta(\tn)=\frac{1}{2(N+M)}\left(
\sum_{i=1}^{N-1} i(N-i)(\tn_i^2-M\tn_i)+2\sum_{i<j}^{N-1} i(N-j) \tn_i\tn_j
\right)\label{dim-formu},
\ea
by using
\ba
\vec\rho\equiv\sum_i \vec\omega_i,\quad
\vec\rho\cdot\vec\rho=\frac{1}{12}N(N^2-1),\quad
\vec\rho\cdot \vec \omega_i=\frac{i(N-i)}{2}\,.
\ea

\subsection{Triality}\label{s:triality}
It is sometimes useful to introduce a parametrization
for the central charge as
\ba
c=(N-1)(1-Q^2 N(N+1)),\quad
Q=\sqrt{\beta}-\frac{1}{\sqrt{\beta}}\,.\label{cN}
\ea
For minimal models, we set $\beta=q/p$.
This formula has an obvious symmetry (duality) in the form:
\ba
\sigma_1: \beta\mapsto \frac{1}{\beta}, \ N\mapsto N
\ea
The level-rank duality symmetry is realized for a specific choice:
\ba
\beta=\frac{N+M}{N}\,.
\ea
We will see later that it is realized by the following transformation, 
\ba
\sigma_2: \beta\mapsto \frac{\beta}{\beta-1}, \ N\mapsto M=(\beta-1) N\,.
\ea
The two transformations $\sigma_{1,2}$ do not commute with each other 
and generate the symmetric group $\mathfrak{S}_3$.  
It is isomorphic to the modular group of four points on two sphere.
We note that such a symmetry exists only for specific choices of the parameter $\beta$.

Such automorphism of algebra is referred as ``triality" in the literature on 
$\cW_\infty[\mu]$ algebra \cite{Gaberdiel:2011wb,Gaberdiel:2012ku}.
It is the automorphism of the universal $\cW$-symmetry but holds for the general
choice of parameters.
In the following, we show that a similar automorphism exists for SH$^c$ 
which is also the universal $\cW$ symmetry in a sense that its rank $N$ representation
agrees with the representation of $\cW_N$ algebra for arbitrary $N$.
Some difference exists, however, since we do not take the large $N$ limit.
%which we expect to be equivalent to $\cW_\infty[\mu]$ with an extra $U(1)$ current algebra symmetry.

\subsection{Correspondence of primary fields}
\label{s:primary}
In this subsection, we explain the duality
of primary fields (their number and conformal dimensions)
of the minimal models with the level-rank duality with $(N, M)$
in details.  Although they are well-known, we need to introduce
a graphical representation which will be useful in later sections.

We start from the number of primary states.
It is given by
\ba\label{npri}
\frac{(N+M-1)!}{N! M!}\,,
\ea
which is symmetric between $N$ and $M$.  
It may be explained as follows.  As in the previous subsection
we introduce $\tilde n_N$ such that $\sum_{i=1}^N \tilde n_i=M$.  By using them we introduce
the number of occupations $\{x_{i}\}_{i=1}^{N}$ in the cyclic group $\mathbb{Z}_{N+M}=\{1,2,\dots, N+M\}$ such that $x_{1}=N+M$;
\ba
x_1=N+M>x_2>\dots >x_N\geq 1.
\ea
$\tilde n_i$ can be derived from such set by
\begin{gather}
\begin{aligned}
x_i-x_{i+1}&= n_i = \tn_i+1, (1\le i\le N-1),\\
x_{N}&=n_{N}=\tn_{N}+1.
\end{aligned}
\end{gather}
The allowed choices of $x_i$ are counted as $\frac{(N+M-1)!}{(N-1)!M!}$.
We will show just after (\ref{confdim}) that
there is an overcounting associated with the rotational symmetry 
$\mathbb{Z}_N$ ($\tilde n_i\mapsto \tilde n_{i+1}$ where
 $ i$ is defined as mod $N$).  We need to divide it by $N$ and it gives (\ref{npri}).

\begin{figure}[]
\centering
\begin{tikzpicture}[baseline=0pt)]
  \draw[thick] (0,0) circle(2);
  \foreach \angle in {1,2,3,6,7}
  \draw[] (0,0) -- (\angle*45+45:2) arc (\angle*45+45:\angle*45+90:2) -- cycle;
  \foreach \angle in {4,5,8}
  \draw[fill=gray!50] (0,0) -- (\angle*45+45:2) arc (\angle*45+45:\angle*45+90:2) -- cycle;
  \foreach \ang in {1,2,...,8}
  \node (\ang) at (\ang*45+45+22.5:1.5) {$\ang$};
  \draw[] (6*45+45:2.2) to[bend right=45] node[right]{$\tn_{1}=2$}(7*45+90:2.2);
  \draw[] (5*45+45:2.1) to (5*45+45:2.4)node[below]{$\tn_{2}=0$};
  \draw[] (1*45+45:2.2) to[out=180,in=135,distance=2cm] node[left]{$\tn_{3}=3$}(3*45+90:2.2);
  \node[above] at (8*45+67.5:2.2) {$x_{1}=8$};
  \node[below right] at (5*45+67.5:2.2) {$x_{2}=5$};
  \node[below left] at (4*45+67.5:2.2) {$x_{3}=4$};
\end{tikzpicture}
\caption{The interpretation of $\tn$.} \label{fig:0}
\end{figure}
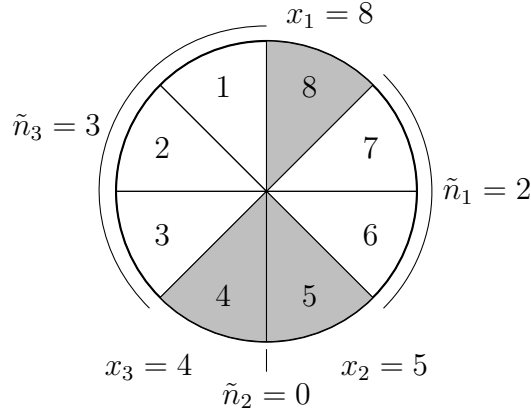

We write a diagram of a disk divided by $N+M$ parts to represent
the $\mathbb{Z}_{N+M}$.  In the diagram we shadow the parts
associated with the set $\{ x_i\}$ (see Figure \ref{fig:0}).
$\tn_i$ is interpreted as the size of
the unoccupied parts between  $x_{i}$ and  $x_{i+1}$ with $x_{N+1}=x_{1}$.
We reorganize the set 
$\{x_{i}\}_{i=1}^{N}$ by identifying the adjacent numbers (in the sense of
$\mathbb{Z}_{N+M}$) as a block.
For instance, we  consider the case $\{x_i\}=\{8,5,4\}$ as 
Figure \ref{fig:0}.
The set is divided into two blocks $\{8\}$ and $\{5,4\}$
and $\tn=(2,0,3)$.
The number of blocks is equal to that of nonvanishing $\tn$
since $\tn_i=0$ implies  $x_{i}$ and $x_{i+1}$ are adjacent. 

We denote the number of blocks by $L\leq \min(N,M)$ in the following. 
Sometimes we relax  the constraint that $N+M$ must be contained in 
$\{x_i\}_{i=1}^{N}$. We get an additional degree of freedom 
to shift all $x_i$'s with a constant number in $\mathbb{Z}_{N+M}$. 
The number of blocks remains the same under such shift.

We define the dual set of numbers $\{y_i\}_{i=1}^{M}$ as the unoccupied set in $\mathbb{Z}_{N+M}$.
Obviously $N+M$ is not an element in this dual set and then we perform a shift in 
$\mathbb{Z}_{N+M}$ to recover the constraint back. 
There are $M$ ways of the shifts which meet the constraint, but they should be identified
in the sense of giving rise to the same sequence 
$(m_i\equiv y_i-y_{i+1})_{i=1}^{M}\subset\mathbb{Z}_{N+M}$ up to $\mathbb{Z}_M$. 
Again we introduce the notation $\tm_i\equiv m_i-1$. 
$\tm_i$ is interpreted as the number of occupied numbers 
between two neighboring unoccupied numbers.
The number of nonvanishing $\tm$ is interpreted as
the number of unoccupied blocks and identical to
$L$ defined above. 
We note here that each nonvanishing $\tm_i$ 
expresses the size of an occupied block,
which is equal to (\# of vanishing $\tn_j$'s)+1 appearing in the block.

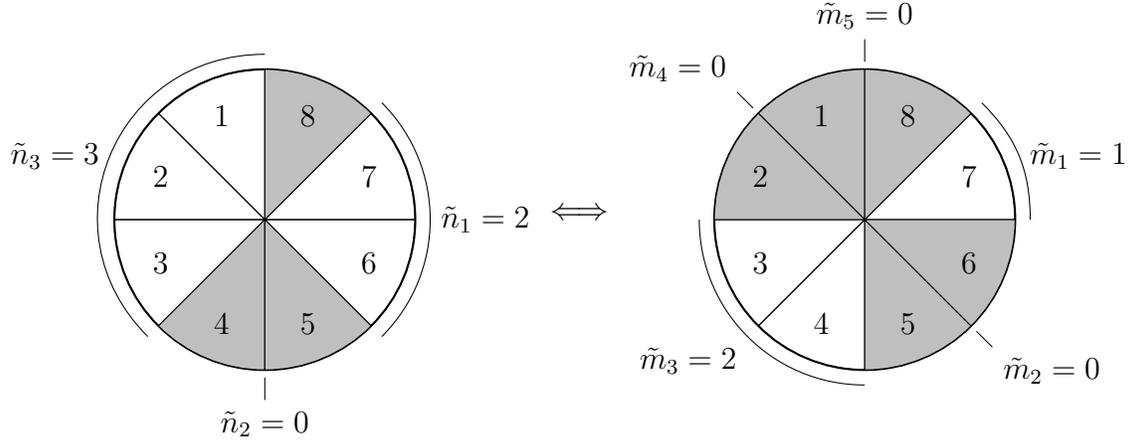
\begin{figure}[]
\centering
\begin{tikzpicture}[baseline=0pt)]
  \draw[thick] (0,0) circle(2);
  \foreach \angle in {1,2,3,6,7}
  \draw[] (0,0) -- (\angle*45+45:2) arc (\angle*45+45:\angle*45+90:2) -- cycle;
  \foreach \angle in {4,5,8}
  \draw[fill=gray!50] (0,0) -- (\angle*45+45:2) arc (\angle*45+45:\angle*45+90:2) -- cycle;
  \foreach \ang in {1,2,...,8}
  \node (\ang) at (\ang*45+45+22.5:1.5) {$\ang$};
  \draw[] (6*45+45:2.2) to[bend right=45] node[right]{$\tn_{1}=2$}(7*45+90:2.2);
  \draw[] (5*45+45:2.1) to (5*45+45:2.4)node[below]{$\tn_{2}=0$};
  \draw[] (1*45+45:2.2) to[out=180,in=135,distance=2cm] node[left]{$\tn_{3}=3$}(3*45+90:2.2);
\end{tikzpicture}
$\Longleftrightarrow$
\begin{tikzpicture}[baseline=0pt]
  \draw[thick] (0,0) circle(2);
  \foreach \angle in {3,4,7}
  \draw[] (0,0) -- (\angle*45+45:2) arc (\angle*45+45:\angle*45+90:2) -- cycle;
  \foreach \angle in {1,2,5,6,8}
  \draw[fill=gray!50] (0,0) -- (\angle*45+45:2) arc (\angle*45+45:\angle*45+90:2) -- cycle;
  \foreach \ang in {1,2,...,8}
  \node (\ang) at (\ang*45+45+22.5:1.5) {$\ang$};
  \draw[] (7*45+45:2.2) to[out=90,in=315] node[right]{$\tm_{1}=1$}(7*45+90:2.2);
  \draw[] (6*45+45:2.1) to (6*45+45:2.4)node[below right]{$\tm_{2}=0$};
  \draw[] (3*45+45:2.2) to[bend right=45] node[below left]{$\tm_{3}=2$}(4*45+90:2.2);
  \draw[] (2*45+45:2.1) to (2*45+45:2.4)node[above left]{$\tm_{4}=0$};
  \draw[] (1*45+45:2.1) to (1*45+45:2.4)node[above]{$\tm_{5}=0$};
\end{tikzpicture}
\caption{A sketch shows how to produce a dual set.} \label{fig:1}
\end{figure}

To illustrate the situation, we take the example $\{x_i\}=\{8,5,4\}$ in Fig.\ref{fig:0} again.
As the dual set, one may choose $y=(8,6,5,2,1)$ and  $\tm=(1,0,2,0,0)$ (see Figure \ref{fig:1})
after a shift. The number of nonvanishing $\tn$'s is $L=2$. 

With this preparation, let us show the conformal dimensions of these two dual pictures are manifestly identical.
From the formula  (\ref{dim-formu}), 
\begin{gather}\label{confdim}
\begin{aligned}
2(N+M)\Delta(\tilde{n})=&-\sum_{i=1}^{N-1}i(N-i)\tilde{n}_i\sum_{j\neq i}^{N}\tilde{n}_j+2\sum_{i<j}^{N-1}i(N-j)\tilde{n}_i\tilde{n}_j\\
=&\sum_{i<j}^N\left(-i(N-i)-j(N-j)+2i(N-j)\right)\tilde{n}_i\tilde{n}_j\\
=&\sum_{i<j}^N(i-j)(N+i-j)\tilde{n}_i\tilde{n}_j\\
=&-\frac{1}{2}\sum_{i=1}^N\sum_{d=1}^{N-1}d(N-d)\tilde{n}_{i}\tilde{n}_{i+d}\,.
\end{aligned}
\end{gather}
From the third line to the last line, we implicitly impose a periodic identification $\tn_{i+N}=\tn_i$ for labels with $\tn$. Here we can see that there exists a $\mathbb{Z}_N$ symmetry as we annouced before: $\Delta(\tn)$ is invariant under $\tn_i\mapsto\tn_{i+1}$. 

\begin{figure}[]
\centering
\begin{tikzpicture}[baseline=0pt)]
  \draw[thick] (0,0) circle(2);
  \foreach \angle in {1,2,3,6,7}
  \draw[] (0,0) -- (\angle*45+45:2) arc (\angle*45+45:\angle*45+90:2) -- cycle;
  \foreach \angle in {4,5,8}
  \draw[fill=gray!50] (0,0) -- (\angle*45+45:2) arc (\angle*45+45:\angle*45+90:2) -- cycle;
  \draw[] (6*45+45:2.2) to[bend right=45] node[right]{$\cecn_{1}=2$}(7*45+90:2.2);
  \draw[] (4*45+45:2.4) to[bend right=45] node[below]{$\cecm_{1}=2$}(5*45+90:2.4);
  \draw[] (1*45+45:2.2) to[out=180,in=135,distance=2cm] node[left]{$\cecn_{2}=3$}(3*45+90:2.2);
  \draw[] (8*45+45:2.4) to[out=135,in=0] node[above right]{$\cecm_{2}=1$}(8*45+90:2.4);
\end{tikzpicture}
\caption{The way we fix the allocation of $\tm_i$'s.} \label{fig:2}
\end{figure}
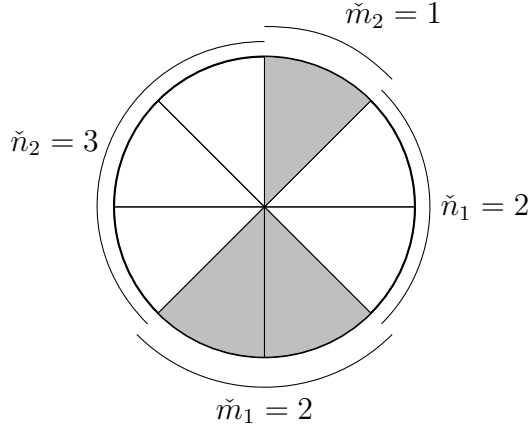

Let $\left( {\check n}_{\check i} \right)_{\ceci =1}^{L}$ denote the subsequence of all nonzero $\tn$'s.
The nonzero contribution in the above sum comes only from two $\cecn_\ceci$'s.
We can see from the interpretation of non-vanising $\tm_{i}$'s noted above that, for such $\cecn_i$'s, the corresponding $d$ and $N-d$ express the sizes of occupied blocks in the two arcs between the two unoccupied blocks, respectively, 
and then can be expressed as summations of some nonvanishing $\tm_i$'s.
In the example of Figure \ref{fig:1}, the nonvanishing $\tilde n_i$ is $\tilde n_1$ and $\tilde n_3$.
The distance parameters are $d=2=\tilde m_3=:\cecm_1$ and $N-d=1=\tilde m_1=:\cecm_2$
 (see Figure \ref{fig:2}).
Now we denote the subsequence of all nonzero $\tm$'s by $(\cecm_\alpha)_{\alpha=1}^{L}$.
We allocate $\tm_i$'s suitably so that
$\cecm_i$ equals the size of the occupied block in the (clockwise) arc 
from the unoccupied block labeled by $\cecn_i$ to the one by $\cecn_{i+1}$. Thus we have
\ba\label{dimdual}
2(N+M)\Delta(\tilde{n})=-\sum_{\ceci<\cecj}\sum_{\ceci\leq\beta<\cecj}\sum_{\cecj\leq\gamma<\ceci+L}\cecm_\beta\cecm_\gamma\cecn_\ceci\cecn_\cecj
=-\sum_{\beta<\gamma}\sum_{\beta<\cecj\leq\gamma}\sum_{\gamma<\ceci\leq\beta+L}\cecn_\ceci\cecn_\cecj\cecm_\beta\cecm_\gamma\,.
\ea
The last expression is manifestly dual by the exchange of $\cecn$ and 
$\cecm$ as $\cecm_{\ceci-1}\mapsto\cecn_\ceci$, $\cecn_\ceci\mapsto\cecm_\ceci$. 
Therefore, the conformal dimension is invariant under the operation 
$(n_{i})_{i=1}^{N}\mapsto (m_{i})_{i=1}^{M}$ 
which represents the duality transformation $\sigma_{2}$.

\subsubsection*{Example: Lee-Yang edge singularity, $(N,M)=(2,3)$}
The simplest example with the duality is given when $(N,M)=(2,3)$.
It will be used repeatedly as the example to illuminate the computation.
We have duality between Virasoro ($N=2$) and $\cW_3$-algebra ($M=3$) 
with the central charge $c=-\frac{22}{5}$.
There are two primary fields.
The diagrams like Figure \ref{fig:1} which may be drawn in $N=2$ (resp. $M=3$) side are (1) $x=(5,4)$ 
(resp. $y=(5,4,3)$)
and (2) $x=(5,3)$ (resp. $y=(5,3,2)$). Corresponding $\tn$ (resp. $\tm$) are $\tn=(0,3)$ (resp. $\tm=(0,0,2)$)
for (1) 
and  $\tn=(1,2)$ (resp. $\tm=(0,1,1)$) for (2).
Their conformal dimensions
are derived from (\ref{dimdual}) as
\ba
\Delta(0,3)=\Delta(0,0,2)=0,\quad
\Delta(1,2)=\Delta(0,1,1)=-\frac15\,.
\ea
%%%%%%%%%%%%%%%%%%%%%%%%%%%%%%%%%%%%%%%%%%

\section{Brief summary of SH$^c$}\label{section3}
The purpose of the paper is to study the level-rank duality and the characterization
of the minimal models in terms of the orthonormal basis
labeled by multiple Young diagrams
and the algebra SH$^c$ which operates on them.
The algebra was explored before in \cite{schiffmann2013cherednik, Kanno:2013aha,Matsuo:2014rba}.
In this section, we summarize some aspects of the algebra and the representation
which will be used in the following sections.
\subsection{Definition of SH$^c$}
The algebra SH$^c$ consists of infinite number of generators 
$D_{r,l}$  with $r\in \mathbb{Z}$ and $l\in \mathbb{Z}_{\geq 0}$.
We refer to the first index $r$ as degree and to the second index $l$ as order.
The commutation relations for degree $\pm 1, 0$ generators
essentially define the algebra,
\begin{gather}
\begin{aligned}\label{SH}
\left[D_{0,l} , D_{1,k} \right] &= D_{1,l+k-1},  &l \geq 1 \,,\\
\left[D_{0,l},D_{-1,k}\right]&=-D_{-1,l+k-1},  &l \geq 1 \,,\\
\left[D_{-1,k},D_{1,l}\right]&=E_{k+l}, & l,k \geq 1\,,\\
\left[D_{0,l} , D_{0,k} \right] &= 0, &k,l\geq 0\,,
\end{aligned}
\end{gather}
where 
$E_l$ is a nonlinear combination of $D_{0,k}$ defined in the form of a generating function,
\begin{gather}
\begin{aligned}\label{com0}
E(\zeta)&=1+(1-\beta)\sum_{l\geq 0}E_l \zeta^{l+1}=\mathcal{C}(\zeta) \mathcal{D}(\zeta),\\
\mathcal{C}(\zeta)&=\exp\left(\sum_{l\geq 0}(-1)^{l+1}c_l \pi_l(\zeta)\right),\\
\mathcal{D}(\zeta)&=\exp\left(\sum_{l\geq 0}D_{0,l+1} \omega_l(\zeta)\right) \,,
\end{aligned}
\end{gather}
with 
\begin{gather}
\begin{aligned}
\pi_l(\zeta)&=\zeta^l G_l\left(1+(1-\beta)\zeta\right) \,,\\
\omega_l(\zeta)&=\sum_{q=1,-\beta,\beta-1}\zeta^l\left(G_l(1-q\zeta)-G_l(1+q\zeta)\right) \,.
\end{aligned}
\end{gather}
The function $G_l(\zeta)$ is defined by the following generating functional,
\ba
\sum_{\ell=0}^\infty G_\ell(\zeta) x^\ell = -\log(\zeta) + \sum_{\ell=1}^\infty x^\ell \frac{\zeta^{-\ell}-1}{\ell}
=\log\left(\frac{1-x}{\zeta-x}\right)\,.
\ea
The other generators $D_{r,l}$ are defined recursively through these commutation relations
repeatedly.

The generators $E_l$ are in general nonlinear combinations of 
$D_{0,k}$ and their commutation relation becomes nonlinear.
The parameters $(c_l)_{l\ge0}$ in the $\cC(\zeta)$
are central charges. They are arbitrary in general.

\subsection{Triality in SH$^c$}
We note here that there exists an $\mathfrak{S}_3$ automorphism
of the algebra SH$^c$ which are related to the triality
in subsection \ref{s:triality}. We repeat to use the same notation for
two generators $\sigma_1, \sigma_2$.
We assume that $\beta\neq 0,1, \infty$ in the following.

The first transformation, $\sigma_1 : \beta\mapsto \beta'=1/\beta$, keeps the algebra invariant if we rescale the parameters and the generators as follows. We first rescale
\ba
\zeta'=-\beta \zeta\,,\quad
c'_\ell = (-\beta)^{-\ell} c_\ell\,,\quad
D'_{0,l+1}=(-\beta)^{-l}D_{0,l+1}\,,
\ea
which make $\cC(\zeta)$ and $\cD(\zeta)$ invariant.
From the definition (\ref{com0}), the generator $E_l$ should be rescaled as,
$
E'_\ell=(-\beta)^{-\ell} E_\ell
$.
The other parts of the algebra (\ref{SH}) are kept invariant by the replacements,
$
D'_{\pm 1,k}=\gamma^{\pm 1}(-\beta)^{-k}D_{\pm 1,k}\,
$ with an arbitrary nonvanishing parameter $\gamma$. 
The other generators are rescaled following the commutation relations.

For the second transformation, $\sigma_2: \beta\mapsto \frac{\beta}{\beta-1}$,
%the transformation of generators and parameters
%is the symmetry of the algebra when all the central charges vanish $c_i=0$,
%namely $\mathcal{C}(\zeta)=1$.
%SH$^c$ without the central extension is called SH.  SH is symmetric under $\sigma_2$.
$\cD(\zeta)$ is invariant if we rescale,
\ba\label{rescale1}
\zeta'=(\beta-1)\zeta, \quad
D'_{0,l+1}=(\beta-1)^{-l}D_{0,l+1}\,.
\ea
We have to tune the central charges by imposing $\cC(\zeta)=\cC'(\zeta')$
which is decomposed into an infinite number of constraints 
by comparing the coefficients of Taylor expansion with respect to $\zeta$.
After taking the logarithm, one obtains,
\ba
\log(\cC'(\zeta'))&=& c'_0\log(1-\zeta)+\sum_{l=1}^\infty
 (-1)^{l+1} c'_l (\beta-1)^l\zeta^l \frac{(1-\zeta)^{-l}-1}{l}\nn\\
 &=& -c'_0 \zeta +\sum_{l=1}^{\infty}\zeta^{l+1}\left( (1-\beta)^l c'_l+ \mbox{(terms written
 	by $c'_0,\cdots,c'_{l-1}$)}
 \right)\\
\log(\cC(\zeta))&=& c_0\log(1+(1-\beta)\zeta)+\sum_{l=1}^\infty
(-1)^{l+1} c_l \zeta^l \frac{(1+(1-\beta)\zeta)^{-l}-1}{l}\nn\\
&=& (1-\beta)c_0 \zeta +(1-\beta)\sum_{l=1}^{\infty}\zeta^{l+1}\left((-1)^l  c_l+ 
\mbox{(terms written
	by $c_0,\cdots,c_{l-1}$)} \right)
\ea
Comparing the coefficient of $\zeta^l$, one can determine $c'_l$ through $c_0,\cdots, c_l$
as $c'_l=(\beta-1)^{-l+1}c_l+$
(terms written by $c_0,\cdots,c_{l-1}$).
Explicit forms of the first three terms are written as,
\ba\label{c-rels}
c_0'=(\beta-1) c_0,\quad
c_1'=c_1+(1-\beta/2)c_0,\quad
c_2'=\frac{-(6-7\beta+2\beta^2)c_0+6(2-\beta)c_1+6c_2}{6(\beta-1)}.
\ea
Finally the other part of the algebra is kept invariant if
\ba\label{rescale2}
E_\ell'=(\beta-1)^{1-\ell}E_\ell,\quad
D'_{\pm 1, \ell}=\gamma^{\pm 1} (\beta-1)^{1/2-\ell}D_{\pm 1, \ell}.
\ea
Again $\gamma$ is an arbitrary nonvanishing number.

In the rank $N$ representation which will be discussed in the next subsection,  
the basis of the representation space is labeled by
the $N$-tuple Young diagrams $\vec Y=(Y_1,\cdots, Y_N)$.
The first transformation $\sigma_1$ is realized
by mapping $\vec Y$  into their transpose $\vec Y'=(Y'_1,\cdots, Y'_N)$.
On the other hand it is in general impossible to express the
second transformation in terms of the finite rank representations
since the number of the parameters in the model is finite.
%which may not be compatible with the invariance of $\cC(\zeta)$.
For specific choices of parameters that correspond to the level-rank duality,
however, $\cC(\zeta)$ is kept invariant in a nontrivial way and
the correspondence between $\vec Y$ becomes less trivial as we will see.

%When $\cC(\zeta)\neq 1$, $\sigma_2$ does not keep it invariant
%in general.  As we will see, however, for specific choices of parameters that correspond to
%the level-rank duality, $\cC$ can be kept invariant.

\subsection{Rank $N$ representation of SH$^c$}\label{SHrep}
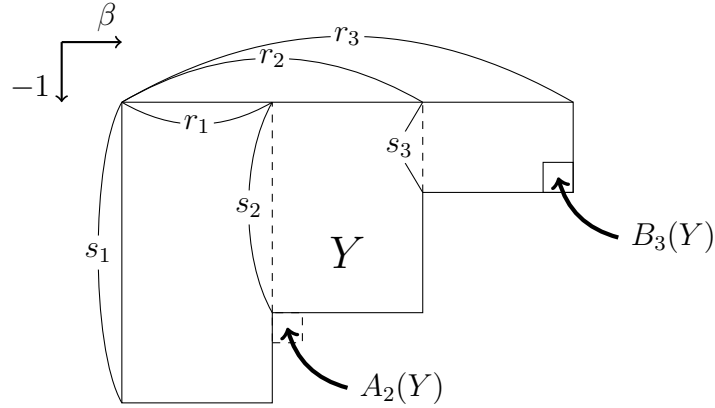
\begin{figure}
\centering
\begin{tikzpicture}[scale=.4]
\coordinate (A)at(0,0);
\coordinate (B)at(0,-10);
\coordinate (C)at(5,-10);
\coordinate (D)at(5,-7);
\coordinate (E)at(10,-7);
\coordinate (F)at(10,-3);
\coordinate (G)at(15,-3);
\coordinate (H)at(15,0);
\node at (7.5,-5) {\Large $Y$};
\draw (A)--(B)--(C)--(D)--(E)--(F)--(G)--(H)--(A);
\draw[->, thick] (-2,2) to node[near end, above]{$\beta$} (0,2);
\draw[->, thick] (-2,2) to node[near end,left]{$-1$} (-2,0);
\draw[dashed] (D)--([yshift=7cm]D);
\draw[dashed] (F)--([yshift=3cm]F);
\draw[bend right, distance=2.1cm] (A) to node [fill=white, inner sep=0.2pt,circle] {$s_{1}$} (B);
\draw[bend right, distance=2.1cm] ([yshift=7cm]D) to node [fill=white, inner sep=0.2pt,circle] {$s_{2}$} (D);
\draw[bend right, distance=2.1cm] ([yshift=3cm]F) to node [fill=white, inner sep=0.2pt,circle] {$s_{3}$} (F);
\draw[bend right] (A) to node [fill=white, inner sep=0.2pt,circle] {$r_{1}$} ([yshift=7cm]D);
\draw[bend left] (A) to node [fill=white, inner sep=0.2pt,circle] {$r_{2}$} ([yshift=3cm]F);
\draw[bend left] (A) to node [fill=white, inner sep=0.2pt,circle] {$r_{3}$} (H);
\draw[dashed] (D) rectangle ([xshift=1cm, yshift=-1cm]D);
\draw [] (G) rectangle ([xshift=-1cm, yshift=1cm]G);
\draw[->,bend left,ultra thick] ([xshift=2.5cm,yshift=-2.5cm]D)node[right]{$A_{2}(Y)$}to([xshift=.5cm,yshift=-.5cm]D);
\draw[->,bend left,ultra thick] ([xshift=1.5cm,yshift=-1.5cm]G)node[right]{$B_{3}(Y)$}to([xshift=-.5cm,yshift=.5cm]G);
\end{tikzpicture}
\caption{Rectangle decomposition of a Young diagram $Y$}\label{rectangledecomp}
\end{figure}

There is a well-studied 
representation where the basis of the
Hilbert space is labeled by $N$-tuple Young diagrams
which is referred as the rank $N$ representation of SH$^c$.
The representation has parameters $\vec a=(a_{q})_{q=1}^{N}\in \mathbb{C}^N$
with which the central charges are written in the form,
\ba
c_l= \sum_{q=1}^N (a_q-\xi)^l\,, \quad \xi\equiv 1-\beta\,.
\ea
With this choice for the central charges, 
the central charge $\mathcal{C}(\zeta)$ is simplified as
\cite{Kanno:2013aha},
\ba
\mathcal{C}(\zeta)=\mathcal{C}_N(\zeta,\vec{a})\equiv\prod_{q=1}^N T(\zeta,a_q),
\quad T(\zeta,a)\equiv\frac{1+\zeta a}{1+\zeta(a-\xi)}\,.
\label{tsa}
\ea
The algebra SH$^c$ is realized on the vector space spanned by the orthogonal
basis $\ket{\vec a,\vec Y}$, where $\vec Y=(Y_q)_{q=1}^{N}$ is an $N$-tuple
Young diagrams.  Strictly speaking $\vec a$ is the parameter of the algebra but
we include it to the state to specify the parameter of algebra which we use.
The highest weight state is given when all the Young diagrams are 
trivial $\vec Y=\vec\emptyset$.  It satisfies the highest weight state conditions,
\ba
D_{-1,l}|\vec a,\vec \emptyset\rrangle=0\,,\quad l=0,1,2,\cdots
\ea 
with $D_{0,l}|\vec a,\vec \emptyset\rrangle=0$.
For $\vec Y\neq \vec \emptyset$, they describe the excited states.
The basis is simultaneous eigenvectors for the infinite number
of generators $D_{0,l}$ $(l\geq 0)$\footnote{
For $N=1$ case, the generators $D_{0,l}$ can be identified with the
infinite number of commuting operators of Calogero-Sutherland system
and the basis is identified with the (normalized) Jack
polynomial $J_Y$.  For the generic $N$, it is identified as a
generalization.
While there seems to exist no explicit proof, the basis $|\vec a,\vec Y\rrangle$
may be identified as the (normalized) basis which was introduced
by \cite{Alba2011, Fateev:2011hq}}.
The other operators $D_{\pm 1,l}$ are defined to change the number of boxes
of the Young diagrams by $\pm 1$ in all possible ways with appropriate coefficients.
The actions of $D_{\pm 1, l}, D_{0,l}$ on the basis are written in a closed form
\cite{schiffmann2013cherednik}\footnote{
This is a generalization of Pieri rule \cite{stanley1989some}
which defines the recursion relation among Jack polynomial.
We use the convention of  \cite{Kanno:2013aha} where the normalized basis was used.},
\ba
D_{-1,l}|\vec a,\vec Y\rrangle&=&(-1)^{l} \sum_{q=1}^{N} \sum_{t=1}^{f_q}(a_q+B_t(Y_q))^l  \Lambda^{(t,-)}_q(\vec Y)\ket{\vec a,\vec Y^{(t,-),q}}\label{D-1} \,,\\
D_{1,l}|\vec a,\vec Y\rrangle&=&(-1)^{l}\sum_{q=1}^{N}\sum_{t=1}^{f_q+1}(a_q+A_t(Y_q))^l  
\Lambda^{(t,+)}_q(\vec Y)\ket{\vec a,\vec Y^{(t,+),q}}\label{D1}\,,\\
D_{0,l+1}|\vec a,\vec Y\rrangle&=&(-1)^l \sum_{q=1}^{N}\sum_{\mu \in Y_q}(a_q+c(\mu))^l \ket{\vec a,\vec Y}\,,
\label{D0}
\ea
where $c(\mu)=\beta i-j$ for $\mu=(i,j)$ describes the coordinates  
($i\geq 1$ represents the horizontal and $j\geq 1$ the vertical location,
$(1,1)$ represents the box at the left-upper corner) of the Young diagram.
The symbols $A_k(Y) = \beta r_{k-1}-s_k-\xi$ and $B_k(Y)=\beta r_k-s_k$ 
(with $r_0=s_{f+1}=0$) parameterize the rectangle decomposition
of Young diagram,
 (see Figure \ref{rectangledecomp}).
The $\vec{Y}^{(t,+),q}$ (resp. the $\vec{Y}^{(t,-),q}$) is the $N$-tuple Young diagrams obtained by adding a box at  
$(r_{t-1}+1, s_{t}+1)$ to (resp. subtracting a box at $(r_{t}, s_{t})$ from) $Y_{q}$ of $\vec{Y}$,
and the coefficients $\Lambda$ are defined as,
\begin{gather}
\begin{aligned}\label{Lambda}
\Lambda^{(k,+)}_p(\vec a,\vec Y) &= \left(
\prod_{q=1}^N \left(\prod_{\ell=1}^{f_q} \frac{
a_p-a_q+A_k(Y_p)-B_\ell(Y_q)+\xi
}{
a_p-a_q+A_k(Y_p)- B_\ell(Y_q)
}{\prod}_{\ell=1}^{\prime f_q +1}\frac{a_p-a_q+A_k(Y_p)- A_\ell(Y_q) -\xi}{
a_p-a_q+A_k(Y_p)-A_\ell(Y_q)
}
\right)\right)^{1/2},
\\
\Lambda^{(k,-)}_p(\vec a,\vec Y) &= \left(
\prod_{q=1}^N \left(\prod_{\ell=1}^{ f_q+1} \frac{ a_p-a_q+B_k(Y_p)-A_\ell(Y_q)-\xi
}{a_p-a_q+B_k(Y_p)- A_\ell(Y_q)
}{\prod}_{\ell=1}^{\prime  f_q}\frac{a_p-a_q+B_k(Y_p)-B_\ell(Y_q)+\xi}{
 a_p-a_q+B_k(Y_p)-B_\ell(Y_q)
}\right)\right)^{1/2},
\end{aligned}
\end{gather}
where the prime in the product symbol ($\prod'$) represents that $(\ell,q)=(k,p)$ is excluded
in the product.  While they are complicated, they play the fundamental role
in the analysis in this paper.

As the action of $D_{0,l}$ on the basis $\ket{\vec a,\vec Y}$ is diagonal,
so is the action of $E_l$.   A nontrivial part is the evaluation 
of the eigenvalue of 
$\mathcal{D}(\zeta)$ which turns out to be written in a compact form,
\begin{gather}
\begin{aligned}
\mathcal{D}(\zeta)\ket{\vec a,\vec Y}
&=\prod_{q=1}^N \Omega(\zeta,a_q, Y_q)\ket{\vec a,\vec Y},\\
\Omega(\zeta,a, Y)&\equiv\prod_{\mu\in Y}\frac{(1+\zeta(a+c(\mu)+1))(1+\zeta(a+c(\mu)-\beta))(1+\zeta(a+c(\mu)-\xi))}
{(1+\zeta(a+c(\mu)-1))(1+\zeta(a+c(\mu)+\beta))(1+\zeta(a+c(\mu)+\xi))}\,.
\end{aligned}
\end{gather}
The factor $\Omega(\zeta,a,Y)$ takes simpler form when a Young diagram $Y$ is divided into rectangles.
The building block is the factor for a rectangle
$Y=(m)^n$ (a rectangle with $m$ rows and $n$ columns),
after some cancellation of factors, we have
\ba
\Omega(\zeta, a, (m)^n)=\frac{T(\zeta,a+n\beta)T(\zeta,a-m)}{T(\zeta,a) T(\zeta,a+n\beta-m)}\,.
\ea
Then we may express it as, with the rectangle decomposition of a Young diagram $Y=(r_t, s_t)_{t=1}^{f}$ $(s_1>\dots>s_f>s_{f+1}=0, 0=r_{0}<r_1<\dots<r_f)$, 
\ba
\Omega(\zeta,a,Y) = \frac{1}{T(\zeta,a)}\frac{\prod_{t=1}^{f+1} T(\zeta, a+A_t(Y)+\xi)}{\prod_{t=1}^f T(\zeta,a+B_t(Y))}\,.
\label{OmegaY}
\ea  
The eigenvalue for the generating functional $E(\zeta)$ is written in a compact form,
\begin{gather}
\begin{aligned}\label{ezeta}
E(\zeta)\ket{\vec a, \vec Y} &= \left(\prod_{p=1}^N \Delta(\zeta,a_p,Y_p)\right)\ket{\vec a, \vec Y},\\
\qquad \Delta(\zeta,a,Y) &=T(\zeta,a) \Omega(\zeta,a, Y)=\frac{\prod_{t=1}^{f+1} T(\zeta, a+A_t(Y)+\xi)}{\prod_{t=1}^f T(\zeta,a+B_t(Y))}.
\end{aligned}
\end{gather}

\section{Description of minimal models in SH$^c$}\label{section4}
In this section, we first explain the translation rule between the representation of $\cW$-algebra
and SH$^c$.
For the minimal model the parameters $\vec a$ in SH$^c$
need to take a special value.
For such choice of parameters, the Hilbert space of SH$^c$ is supposed to
be the same as that of $\cW_N$ with an extra $\mathrm{U}(1)$ factor
where the extra part is realized as the Fock space spanned by one free boson.
We show that this is indeed the case by studying the structure of null states
in the Hilbert space are identical.  We will also come back to the similar setup
in section \ref{section7} and provide other characterization of $\cW$ module, $N$-Burge
condition through SH$^c$.
%%%%%%%%%%%%%%%%%%%%%%%%%%%%%%
\subsection{Correspondence between parameters in SH$^c$ and minimal models}
The representation of SH$^c$ in
(\ref{SHrep}) is parametrized by the central charges,
which is written in terms of the vector $\vec a\in \mathbb{C}^N$. 
On the other hand, the minimal model of $\cW_N$-algebra
is parametrized by a sequence of integers, 
which in general labels  
the momentum of scalar fields in Toda field theory.
The vertex operator with a momentum $\vec{p}$ in the conformal Toda theory has the conformal dimension,
\ba
\Delta(\vec p) =\frac{\vec p\cdot(\vec p-2Q\vec\rho)}{2}.
\ea
Comparing with the weight formula (\ref{weight}), we may express $\vec p$ by $(n, n')$ as follows,
\ba
\vec p=\sum_{i=1}^{N-1} \left(
\beta^{-1/2} n_i -\beta^{1/2} n'_i
\right)\vec \omega_i+Q\vec \rho=\sum_{i=1}^{N-1} \left(
\beta^{-1/2} \tilde n_i -\beta^{1/2} \tilde n'_i
\right)\vec \omega_i\,,
\ea
with $\beta=q/p$.
The parameter $\vec a$ in the representation of SH$^c$ is related to $\vec p$
as \cite{Kanno:2013aha}
\ba
a_{i}=-\sqrt{\beta}p_{i}+i\xi.
\ea

We note that $\vec a$ and $\vec a^{\epsilon}:=(a_{\epsilon(1)},\cdots,a_{\epsilon(N)})$
gives the same representation of SH$^c$ for $\epsilon\in\mathfrak{S}_{N}$
since the structure constants are symmetric.
So one may set $\vec{a}$ by $a_{\epsilon(i)}=-\sqrt{\beta}p_{i}+i\xi$ . 
Since the inner product (\ref{omegaproduct}) is symmetric under $i\leftrightarrow N-i$, we have another way to connect the momentum and $(n,n')$, given by
\begin{gather}
\vec{p'}=\sum_{i=1}^{N-1} \left(
\beta^{-1/2} \tilde n_{N-i} -\beta^{1/2} \tilde n'_{N-i}
\right)\vec \omega_{i},
\end{gather}
In the following sometimes the choice $\epsilon=(N, N-1,\cdots, 1)\in \mathfrak{S}_N$
will be useful  and we may set
\begin{gather}\label{aitonprime}
a'_{N-i}=-\sqrt{\beta}p'_{i}+i\xi,
\end{gather}
with $a'_{0}=a'_{N}$. This identification will appear in the next section. Note again that the order of indices in $\vec{a}'$ is not important.

For the particular representations which appear in the level-rank duality,
we have $\tilde n'_i=0$ for $1\le i\le N-1$.  Also, one should put the parameters as
\ba
\beta=\frac{N+M}{N}\,,\quad
Q=\frac{M}{\sqrt{(N+M)N}}=\sqrt{\frac{M+N}{N}}-\sqrt{\frac{N}{N+M}},\quad
\xi = -\frac{M}{N}\,.
\ea
Then some formulae are simplified, and we have
\ba\label{aiton}
\vec a=-\sum_{i=1}^{N-1} \tn_i\vec\omega_i+\xi \vec I\,,
\ea
where $\vec{I}=(1,2,\dots,N)$.

\subsection{Singular vectors of minimal models in SH$^c$}
As an application of the correspondence between the
parameters in $\cW_N$-algebra
and SH$^c$, let us make a comparison between the structure
of null states in the Hilbert space for the general minimal models.

In the case of $\cW_N$-algebra, in the highest weight module
with (\ref{weight}), there are $N$ null states at level $n_i n'_i$
($1\le i\le N$). We show that exactly the same null states appear in the SH$^c$ algebra.
The null states are characterized by the condition,
\ba\label{null}
D_{-1,\ell}|\vec a, \vec Y\rangle=0,\quad\ell\ge0.
\ea
This condition can be satisfied when
the factor $\Lambda^{(k,-)}_p(\vec a,\vec Y)$ in (\ref{Lambda}) vanishes for arbitrary $p,k$.  It occurs when at least one numerator factor
in (\ref{Lambda}) vanishes for any $p,k$.

To see null states, first we determine differences among $(a_{q})_{q=1}^{N}$. Let $(\vec e_i)_{i=1}^{N}$ be the standard basis of $\mathbb{R}^N$.
The fundamental weight is determined by the relation $\vec \omega_i\cdot \vec\alpha_j=\delta_{i,j}$
with $\vec\alpha_i=\vec e_{i}-\vec e_{i+1}$. 
We find that
\begin{gather}
\begin{aligned}\label{neighbor-a}
a_{j+1}-a_j&=-(\vec \alpha_j, \vec a)=n_j-\beta n'_j, \quad (1\le j\le N-1),\\
a_1-a_N &= n_N-\beta n'_N\,,
\end{aligned}
\end{gather}
where we used $n_N=N+M-\sum_{i=1}^{N-1}n_i$ and $n'_N\equiv N-\sum_{i=1}^{N-1}n'_i$. 

We consider the case when $Y_q=\emptyset$ ($q\neq j+1$)
and $Y_{j+1}= (n_j)^{n'_j}$ (a rectangle Young diagram).
For this case, there is only one box which can be removed with the action of $D_{-1,l}$ and we have $A_1(\emptyset)+\xi=0$ and $B_1((n_j)^{n'_j})=\beta n'_j-n_j$.
Then a factor $a_{j+1}-a_{j}+B_1(Y_{j+1})-A_{1}(Y_{j})-\xi$ in $\Lambda^{(1,-)}_{j+1}$ vanishes.  
It implies that the condition
(\ref{null}) is satisfied for $N$ states specified by,
\ba
\{\vec Y\}=\{(n_N)^{n'_N},\emptyset,\dots,\emptyset),
(\emptyset,(n_1)^{n'_1},\emptyset, \dots,\emptyset),\dots,
(\emptyset,\dots,\emptyset,(n_{N-1})^{n'_{N-1}})\}.
\ea
The levels of these states are $n_i n'_i$ ($1\le i\le N$)
which coincide with the null states in $\cW_N$-algebra.\footnote{
In \cite{mimachi1995singular, awata1995excited}, the singular vectors
in the Virasoro module are identified with the Jack polynomials that
correspond to the rectangle Young diagrams.  Our statement here
is a generalization of such observation since AFLT basis may be
regarded as a generalization of Jack polynomials \cite{Alba2011}.}
This observation agrees with our expectation that the representation of SH$^c$
with an $N$-tuple Young diagrams is identical to the representation of $\cW_N$-algebra
with the extra $U(1)$ factor which is not relevant in the degeneracy of the Hilbert space.

These null states
may be interpreted as the highest weight states with the parameter
$(\vec n, \vec n')$ shifted.  Let us focus on the case where $Y_{j+1}=(n_j)^{n'_j}$
and other $Y_q$'s are $\emptyset$.  
This state has the eigenvalue of $E(\zeta)$ which (\ref{ezeta}):
\begin{gather}
\begin{aligned}
&\left(\prod_{q\neq j+1} T(\zeta, a_q) \right)\frac{T(\zeta, a_{j+1}+n_j'\beta)T(\zeta, a_{j+1}-n_j)}{
T(\zeta, a_{j+1}+n'_j \beta-n_j)}\\
&=\left(\prod_{q\neq j,j+1} T(\zeta, a_q) \right)
T(\zeta, a_{j+1}-n_j) T(\zeta, a_j+n_j) = \cC_N(\zeta,\vec b)\,.
\end{aligned}
\end{gather}
What is interesting here is that the eigenvalue is written as
a product of $N$ factors of the form of  $\cC_N(\zeta, \vec b)$
with,
\ba
b_q=\left\{
\begin{array}{ll}
a_{j+1}-n_j \quad & q=j+1\\
a_j+n_j\quad & q=j\\
a_q \quad &\mbox{otherwise.}
\end{array}
\right. 
\ea
So the eigenvalues of the excited state are identical to those of ground state
after the parameter shift.

We note that the singular vectors are constructed through the
screening operators in the $\cW_N$ algebra.  In SH$^c$, on the other hand,
they are already built in the coefficients of the representation
(\ref{Lambda}). 

\section{Level-rank duality in SH$^c$: central charges}\label{section5}
We come back to the level-rank duality in the SH$^c$ algebra. 
The focus of this section is to prove the central charges $\cC(\zeta)$ in SH$^c$ are
identical for the dual pair.  We have seen in (\ref{rescale1},\ref{rescale2}) that
the other parts of the algebra are identical after suitable
rescaling of parameters and generators.
On the other hand, $\cC(\zeta)$ is kept invariant by the tuning infinite parameters $c_n$.
In the rank $N$ representation, they are replaced by finite parameters $a_p$
and it is not obvious if $\cC(\zeta)$ can be kept invariant.
%This is not so obvious since two models contains different number of parameters with which
%the center of algebra is written in the form of generating functional (\ref{tsa}).
If they are identical, it establishes the triality of SH$^c$  for the minimal model CFTs.

The main statement of the section is the following; {\it with the correspondence between $\tn$ and $\tm$ appeared in subsection \ref{s:primary}, 
the central charges of SH$^c$ are identical after rescaling (\ref{rescale1}):}
\ba\label{TT}
\cC_N(\zeta, \vec a)=\prod_{q=1}^N T(\zeta, a_q) =\prod_{q=1}^M T(\zeta', a'_q)
=\cC_M(\zeta', \vec a').
\ea
We have to be careful in the definition of $a'$ in the dual pair
which will be explained in the proof  (\ref{explicita2}). 

\subsection*{Proof}
First we prepare the explicit form of $a_q$ and $a_q'$ in the main statement (\ref{TT}).
We recall that, with $\tn_N=M-\sum_{i=1}^{N-1}\tn_i$, we rewrite (\ref{aiton}) as,
\begin{gather}
\begin{aligned}\label{explicita1}
a_i&=\sum_{j<i}\frac{j}{N}\tilde{n}_j-\sum_{j\geq i}^{N-1}\frac{N-j}{N}\tilde{n}_j-\frac{iM}{N}\\
&=\sum_{j<i}\frac{j-i}{N}\tilde{n}_j-\sum_{j\ge i}^{N}\frac{N-(j-i)}{N}\tilde{n}_j.
\end{aligned}
\end{gather}
Similarly, we rewrite (\ref{aitonprime}) and get
\begin{gather}
\begin{aligned}\label{explicita2}
a'_{i}&=\sum_{j>i}^{N}\frac{i-j}{N}\tilde{n}_{j}-\sum_{j\le i}\frac{N-(i-j)}{N}\tilde{n}_{j}.
\end{aligned}
\end{gather}
Then we have
\begin{gather}
\left\{
\begin{aligned}\label{adifference}
a_{j+1}-a_j&=\tn_j+\xi, \\
a_{1}-a_{N}&=\tn_{N}+\xi,
\end{aligned}
\right.\\
\left\{
\begin{aligned}
a'_{j}-a'_{j-1}&=-\tn_{j}-\xi, \\
a'_{1}-a'_{N}&=-\tn_{1}-\xi.
\end{aligned}\right.
\end{gather}

Here we reuse the notation in subsection \ref{s:primary}. 
We denote by $(\cecn_\ceci)_{\ceci=1}^{L}$ the subsequence of all nonvanishing $\tn_i$'s as well as by $(\cecm_\ceci)_{\ceci=1}^{L}$, allocated so that the block with $\cecm_\ceci$ appears just after the one with $\cecn_\ceci$ in the ``clock" (see Figure \ref{fig:2} in subsection \ref{s:primary}). We denote by $j(\ceci)$ the labeling satisfying $\cecn_{\ceci}=\tn_{j(\ceci)}$ and that the block with $\tn_{j(\ceci+1)}$ is the next unoccupied block to the one with $\tn_{j(\ceci)}$. We also denote by $k(\ceci)$ for $(\cecm_\ceci)_{\ceci=1}^{L}$, which satisfies $\cecm_{\ceci}=\tm_{k(\ceci)}$.

Recall that $a_{j+1}-\xi=a_{j}$ when $\tn_{j}=0$, and then we have
\ba
\mathcal{C}_{N}(\zeta, \vec a)=\prod_{\ceci=1}^{L}\frac{1+\zeta a_{j(\ceci)}}{1+\zeta(a_{j(\ceci)+1}-\xi)}=\prod_{\ceci=1}^{L}\frac{1+\zeta a_{j(\ceci)}}{1+\zeta(a_{j(\ceci)}+\cecn_{\ceci})}.
\ea
Also we have
\ba
\mathcal{C}_{M}(\zeta', \vec a')=\prod_{\ceci=1}^{L}\frac{1+\zeta' a'_{k(\ceci)}}{1+\zeta'(a'_{k(\ceci)-1}-\xi)}=\prod_{\ceci=1}^{L}\frac{1+\zeta' a'_{k(\ceci)}}{1+\zeta'(a'_{k(\ceci)}+\cecm_{\ceci})}.
\ea
A similar discussion in subsection \ref{s:primary} leads to 
\begin{gather}
\begin{aligned}\label{neq0}
Na_{j(\ceci)}&=-\sum_{\cecj<\ceci}\sum_{\cecj\le\ceck<\ceci}\cecm_{\ceck}\cecn_{\cecj}-\sum_{\cecj\ge\ceci}\sum_{\cecj\le\ceck<\ceci+L}\cecm_{\ceck}\cecn_{\cecj}\\
&=-\sum_{\ceck<\ceci}\sum_{\ceci\le\cecj\le\ceck+L}\cecm_{\ceck}\cecn_{\cecj}-\sum_{\ceck\ge\ceci}\sum_{\ceci\le\cecj\le\ceck}\cecm_{\ceck}\cecn_{\cecj},\\
Ma'_{k(\ceci)}&=-\sum_{\ceck>\ceci}\sum_{\ceci<\cecj\le\ceck}\cecm_{\ceck}\cecn_{\cecj}-\sum_{\ceck\le\ceci}\sum_{\ceci<\cecj\le\ceck+L}\cecm_{\ceck}\cecn_{\cecj}.
\end{aligned}
\end{gather}
Therefore
\begin{gather}
Na_{j(\ceci+1)}=Ma'_{k(\ceci)}.
\end{gather}
Note that
\begin{gather}
Na_{j(\ceci+1)}-Na_{j(\ceci)}=N\cecn_{\ceci}-M\cecm_{\ceci},%\\
%Ma'_{k(\ceci)}-Ma'_{k(\ceci-1)}=M\cecm_{\ceci}-N\cecn_{\ceci},
\end{gather}
then
\begin{gather}\label{aisaprime}
M(a'_{k(\ceci)}+\cecm_{\ceci})=N(a_{j(\ceci)}+\cecn_{\ceci}).
\end{gather}
As a result, we have $\mathcal{C}_{N}(\zeta, \vec a)=\mathcal{C}_{M}(\zeta', \vec a')$. \hfill$\Box$

\subsection*{Example}
To illustrate the statement of the section, we demonstrate the
explicit form of $\mathcal{C}_N(\zeta)$ for $(N,M)=(2,3)$.

For the $\Delta=0$ case, on the $N=2$ side, we take $\tilde{n}=(0,3)$ and then we have
\ba
\{2 \vec a_i\}=\{-3,-6\}\,, \quad \{2(a_i-\xi)\}=\{0,-3\}\,.
\ea
Therefore, after the cancellation between the numerator and denominator, we have
\ba
{\cC}_2(\zeta,\vec a)=1-6\cdot\frac{\zeta}{2}\,.
\ea
On the $M=3$ side, the same state is described by $\tilde{m}=(0,0,2)$. We obtain\footnote{Here we write elements in the decreasing order. Recall that there is an arbitrariness of arrangement of indices in $a_{i}, a'_{j}$.},
\ba
\{3a'_i\}=\{-2,-4,-6\},\quad \{3(a'_i-\xi)\}=\{0,-2,-4\}\,.
\ea
There is again a similar cancellation and we obtain
\ba
{\cC}_3(\zeta',\vec a')=1-6\cdot\frac{\zeta'}{3}\,.
\ea
After the identification $\zeta/2=\zeta'/3$, we obtain (\ref{TT}).

For the $\Delta=-1/5$ case, on the $N=2$ side, we need to take $\tilde{n}=(1,2)$,
which implies $\{2a_i\}=\{-4,-5\}$, $\{2(a_i-\xi)\}=\{-1,-2\}$ and
\ba
{\cC}_2(\zeta, \vec a)=\frac{(1-4\zeta/2)(1-5\zeta/2)}{(1-\zeta/2)(1-2\zeta/2)}.
\ea
In the dual picture, $\tilde{m}=(0,1,1)$ and
 $\{3a'_i\}=\{-3,-4,-5\}$ and $\{3(a'_i-\xi)\}=\{-1,-2,-3\}$.
Again after the cancellation we obtain
\ba
{\cC}_3(\zeta',\vec a') =\frac{(1-4\zeta'/3)(1-5\zeta'/3)}{(1-\zeta'/3)(1-2\zeta'/3)}.
\ea
In either case some cancellations 
between the numerator and denominator occur and the numbers of 
factors match in the end while the dimension of $\vec a$ is different.

\section{Level-rank duality in SH$^c$: state-to-state correspondence}\label{section6}
We have proved that the central charges of SH$^c$ are identical
for the pair of the minimal models with the level-rank duality.
If the two algebras are identical, so should be the representations.
It remains to be curious, however, how the states labeled by
different sets of Young diagrams can be identified.
Of course, it is possible only when some of the basis become null.

We note that in order to identify two states labeled by different sets of Young diagrams,
the eigenvalues of $D_{0,l}$ should match.  Eq.(\ref{D0}) implies that the eigenvalue
is the power sum of the numbers $a_q+c(\mu)$ assigned to each box in $\vec Y$.
The transformation of $D_{0,l}$ under $\sigma_2$ (\ref{rescale1}) becomes,
$D_{0,l+1}'=(N/M)^l D_{0,l+1}$.  By combining them, we need to require 
the rescaled numbers assigned to each boxes in $\vec Y$ and $\vec Y'$,
\ba\label{idset}
\sqcup_{q=1}^N \{N(a_q+c(\mu))\}_{\mu\in Y_q} =
\sqcup_{q=1}^M \{M(a'_q+c(\mu'))\}_{\mu'\in Y'_q}
\ea
to be equal as a set 
in order to identify $|\vec a, \vec Y\rrangle$ and $|\vec a', \vec Y'\rrangle$.  
We refer to $N(a_q+c(\mu))$ as the {\it characterizing number} of the box.
With the help of such observations,
the level-rank duality is realized 
as an intertwining map which shuffles rows of Young diagrams appropriately.
In order to have such consistent identification of boxes in $\vec Y$ and $\vec Y'$,
we take the following steps. 

\paragraph{Step 1 : labels of the rows}
We define the set of the characterizing numbers of the leftmost boxes on each Young
diagrams as X,
\ba
X&=&\sqcup_{q=1}^N\sqcup_{j=1}^\infty\{ N(a_q+c(1,j))\}=
\sqcup_{q=1}^N\sqcup_{j=1}^\infty \{ N a_q+M-N(j-1)\}\nonumber\\
X'&=&\sqcup_{q=1}^M\sqcup_{j=1}^\infty\{ M(a'_q+c(1,j))\}
=\sqcup_{q=1}^N\sqcup_{j=1}^\infty \{ M a'_q+N-M(j-1)\}.
\ea
We claim that they have the following properties:
\begin{itemize}
\item[(1-1)] In either set $X$ or $X'$, there is no overlap of elements. 
\item[(1-2)] If $x\in X$ (resp. $X'$) then $x-N, x-M \in X$ (resp. $X'$).
\item[(1-3)] Two sets are identical: $X=X'$.
\end{itemize}
These statements are nontrivial and will be proved in subsection \ref{s:6.1}.
They imply that there is a one-to-one correspondence between
the leftmost boxes in $\vec Y$ and $\vec Y'$.
We use the number $x\in X$ to label the rows of Young diagrams.

\paragraph{Step 2: Correspondence between states}
Once we have the identification of leftmost boxes, one may assign the
correspondence between the 
boxes in the same rows at the same time.  Suppose $(1,j)\in Y_q$
and $(1,j')\in Y'_{q'}$ to have the same label $x$.  Then both
boxes, $(i,j)\in Y_q$ and $(i,j')\in Y'_{q'}$ have the same characterizing
number $x+(N+M)(i-1)$.  We establish the correspondence between
the rows with the same label
since they do not violate (\ref{idset}).

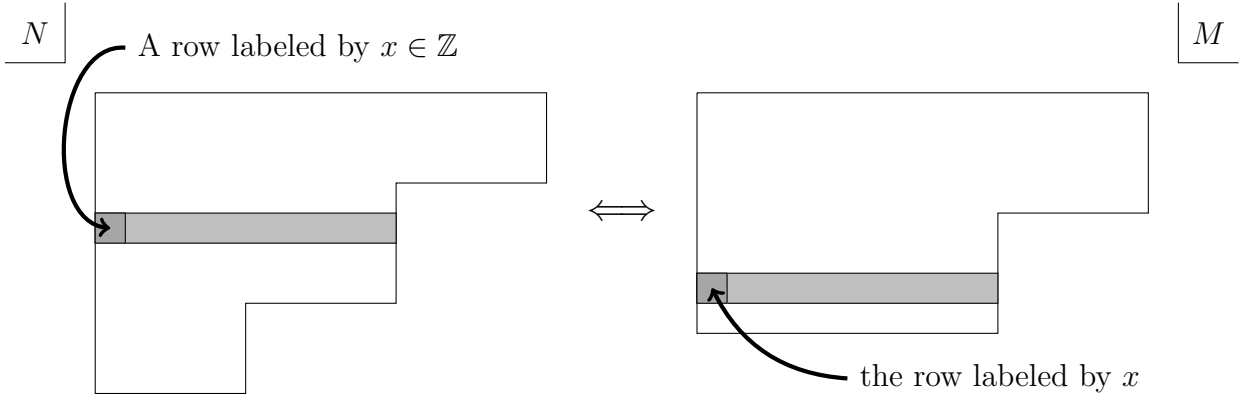
\begin{figure}
\centering
\begin{tikzpicture}[scale=.4]
\coordinate (A)at(0,0);
\coordinate (B)at(0,-10);
\coordinate (C)at(5,-10);
\coordinate (D)at(5,-7);
\coordinate (E)at(10,-7);
\coordinate (F)at(10,-3);
\coordinate (G)at(15,-3);
\coordinate (H)at(15,0);
\node[] at (-2,2) {$N$};
\draw (A)--(B)--(C)--(D)--(E)--(F)--(G)--(H)--(A);
\draw[fill=gray!50] (0,-4) rectangle (10,-5);
\draw[fill=gray!70] (0,-4) rectangle (1,-5);
\draw (-3,1)--(-1,1)--(-1,3);
\draw[->,bend right,ultra thick, in=-90, out=-90] ([yshift=2.5cm]1,-1)node[right]{A row labeled by $x\in\mathbb{Z}$}to([xshift=.5cm,yshift=-.5cm]0,-4);
\node[] at (17.5, -4) {\large$\Longleftrightarrow$};
\coordinate (A1)at(20,0);
\coordinate (B1)at(20,-8);
\coordinate (C1)at(30,-8);
\coordinate (D1)at(30,-4);
\coordinate (E1)at(35,-4);
\coordinate (F1)at(35,0);
\node[] at (37,2) {$M$};
\draw (A1)--(B1)--(C1)--(D1)--(E1)--(F1)--(A1);
\draw[fill=gray!50] (20,-6) rectangle (30,-7);
\draw[fill=gray!70] (20,-6) rectangle (21,-7);
\draw (36,3)--(36,1)--(38,1);
\draw[->,bend left,ultra thick] ([yshift=.5cm]25,-10)node[right]{the row labeled by $x$}to([xshift=.5cm,yshift=-.5cm]20,-6);
\end{tikzpicture}
\caption{Rearrangement of Young diagrams with the level-rank duality. Two gray rows have the same number of boxes.}\label{linecor}
\end{figure}
It implies that
the two states $|\vec a, \vec Y\rrangle$
and $|\vec a', \vec Y'\rrangle$  are
identified if the length of rows with the same label are identical
 (see Figure \ref{linecor}). 
 In general, two rows which belong to the same $Y_q$ may not correspond to two rows
in the same $Y'_{q'}$ in $\vec Y'$.  We call such restructuring of Young diagrams as
the {\it shuffling of rows}.

For the consistency of such identification, the shapes of Young diagrams in $\vec Y$ or $\vec Y'$ 
are constrained.
In order that each $Y_q$ is the Young diagram,
we need to have an inequality $\lambda_{\vec{Y}}(x-N)\le\lambda_{\vec Y}(x)$ for each $x\in X$
where $\lambda_{\vec{Y}}(x)$ denotes the number of boxes in the row characterized 
by the integer $x\in X$ with respect to $\vec{Y}$. 
The mapped shapes form a vector of Young diagrams on the dual side 
if and only if $\lambda_{\vec Y}(x-M)\le\lambda_{\vec{Y}}(x)$ is satisfied for each $x\in X$.

Finally we arrive at the main statement of the section:
\begin{itemize}
\item[(2)] For the dual pair of the minimal models,
the excited states labeled by either
$\vec Y$ or  $\vec Y'$ are obtained from
a single set of the integer partitions $\lambda: X\to \mathbb{N}$, $|\lambda|\equiv\sum_{x\in X}\lambda(x)<\infty$, satisfying the condition 
$\lambda(x-N), \lambda(x-M)\le\lambda(x)$ for each $x\in X$.
Then the shuffling of rows gives an automorphism between the two representations.
\end{itemize}
Drawing an analogy with Young diagrams, we introduce a partial order $<_{X}$ to the set $X$ by
\ba\label{poset}
x\ge_{X}y \Leftrightarrow {}^{\exists} i_{N}, i_{M}\in\mathbb{N},\ x=y-Ni_{N}-Mi_{M}.
\ea
Then we rewrite the condition for the partition for the partially ordered set (poset) 
by
\ba\label{Xpartition}
x\ge_{X}y \Rightarrow \lambda_{\vec Y}(x)\le\lambda_{\vec Y}(y).
\ea

In subsection \ref{s:6.2}, we will show that the module obtained
by applying $D_{1,l}$ to the vacuum state 
$\ket{\emptyset,\dots,\emptyset}$ are spanned by all such restricted states.
However we should recall that we have taken $a_{q}$'s to specific values
for the minimal models. It may cause divergence of $\Lambda^{\pm}$ in (\ref{Lambda}).
It triggers us to check that the highest weight representation is well-defined for the case of the level-rank duality.
We will %also                                                                                                                                          
check it by deriving the condition when the factors in the product in (\ref{Lambda}) vanish.
This will in turn lead to the following statement:
\begin{itemize}
\item[(3)] The highest weight representation is not only well-defined but also is spanned by {\it all} the integer partition over $X$ satisfying (\ref{Xpartition}).
\end{itemize}
Then the claim (2) will follow as a corollary of the statement (3).

\subsection{Example}\label{s:6.5}
Since the main statements of this section may be abstract, it is illuminating
to describe them in concrete examples.  The case $(N,M)=(2,3)$ is the easiest nontrivial example.
We have two primary fields with conformal weights $\Delta=0$ and $\Delta=-1/5$.

For $\Delta=0$ case, we have $\vec a=\{-3/2, -3\}$ and $\vec a'=\{-2/3, -4/3, -2\}$.
The set $X$, $X'$ of labels are computed as,
\ba
X=X'=\{0,-2,-3,-4,-5, -6,\cdots\}\,.
\ea
For $\Delta=1/5$, similar computation gives,
\ba
X=X'=\{-1,-2,-3,-4,-5, -6,\cdots\}\,.
\ea
In either case, $X, X'$ satisfy the properties claimed in step 1.

Let us examine what states are generated by applying $D_{1,l}$ to the vacuum.
We focus on the one with $\Delta=-1/5$ here. 
We use the parameter $\vec{a}=(-2, -5/2)\leftrightarrow\vec{a}'=(-4/3,-1,-5/3)$
but omit them in the label of excited states.
The labels for the Young diagrams (recall Figure \ref{shuffle:intro}) are
\ba
N=2&& Y_1\leftrightarrow \{-1,-3,-5,\cdots\},\quad Y_2\leftrightarrow \{-2,-4,-6,\cdots\}\\
M=3&& Y'_1\leftrightarrow \{-2,-5,-8,\cdots\},\quad Y'_2\leftrightarrow \{-1,-4,-7,\cdots\},\quad
Y'_3\leftrightarrow \{-3,-6,-9,\cdots\}
\ea

At level one (one Young box),
\begin{gather}
\begin{aligned}
D_{1,l}\ket{\emptyset,\emptyset}&=(-1)^l\left(\left(-\frac{1}{2}\right)^l\times2\ket{\ydiagram{1},\emptyset}+(-1)^l\times i\sqrt{2}\ket{\emptyset,\ydiagram{1}}\right)\,,\\
D_{1,l}\ket{\emptyset,\emptyset,\emptyset}&=(-1)^l\left(\left(-\frac{2}{3}\right)^l\times i\sqrt{3}\ket{\ydiagram{1},\emptyset,\emptyset}+\left(-\frac{1}{3}\right)^l\times\sqrt{6}\ket{\emptyset,\ydiagram{1},\emptyset}\right).
\end{aligned}
\end{gather}
After required rescaling $D_{1,l}\mapsto\left(\frac{2}{3}\right)^{l-1/2}D_{1,l}$, 
we can identify,
\ba
\ket{\ydiagram{1},\emptyset}\leftrightarrow\ket{\emptyset,\ydiagram{1},\emptyset}\qquad
\ket{\emptyset,\ydiagram{1})}\leftrightarrow\ket{\ydiagram{1},\emptyset,\emptyset}
\ea
The two states correspond to the partition $\lambda(-1)=1, \lambda(-2)=\lambda(-3)=\cdots=0$
and $\lambda(-2)=1,  \lambda(-1)=\lambda(-3)=\lambda(-4)=\cdots=0$.
We note that a state $\ket{\emptyset,\emptyset, \ydiagram{1}}$ is not created.
It corresponds to a partition $\lambda(-3)=1,  \lambda(-1)=\lambda(-2)=\lambda(-4)=\cdots=0$
but it breaks the rule (\ref{Xpartition}) since $\lambda(-1)<\lambda(-3)$.

We can perform similar  such computation to higher levels in the same way. The correspondence between non-null states in the Lee-Yang case will be explicitly given up to level $5$ in Appendix \ref{s:A}. 
It is easy to see the shuffing rule of rows and the condition (\ref{Xpartition}) are satsified
in all of them.

\subsection{Proof of Step 1}\label{s:6.1}
Here we give a proof of the statements in Step 1.

We first confirm that there is no overlap of elements in
the set, 
\ba\label{defX}
X=\{N(a_{q}+\mu(1,s))| q=1,\dots,N, s\ge1\}
\ea
namely the statement (1-1).
Actually this is obvious since we have $N(a_{q}+\mu(1,s))=Na_{q}+M-N(s-1)$ and 
we see from (\ref{adifference}) that no two integers $Na_{q}$'s modulo $N$ are the same. 

Next we prove the statement (1-2),
\begin{gather}
x\in X \Rightarrow x-N, x-M\in X.
\end{gather}
It is sufficient to show that for $x\in\{Na_{q}+M\}_{q=1}^{N}$, $x+M\in X$. Now we rewrite the subset as
\begin{gather}
\begin{aligned}
\{Na_{q}+M\}_{q=1}^{N}&=\sqcup_{i=1}^{L}\{N(a_{j(\ceci)}+\cecn_\ceci)-M(l-1)\ |\ 1\le l\le \cecm_\ceci\}, \\
\{Ma'_{q}+N\}_{q=1}^{M}&=\sqcup_{i=1}^{L}\{M(a'_{k(\ceci)}+\cecm_\ceci)-N(l-1)\ |\ 1\le l\le \cecn_\ceci\}
\end{aligned}
\end{gather}
by use of the notations appeared in the previous section. Then the following equations shows the property;
\begin{gather}
\begin{aligned}
(N(a_{j(\ceci)}+\cecn_\ceci)-M(\cecm_\ceci-1))-M&= Na_{j(\ceci+1)}= N(a_{j(\ceci+1)}+\cecn_{\ceci+1})-N\cecn_{\ceci+1}\in X,\\
(M(a'_{k(\ceci)}+\cecm_\ceci)-N(\cecn_\ceci-1))-N&=Ma'_{k(\ceci-1)}= M(a'_{k(\ceci-1)}+\cecm_{\ceci-1})-M\cecm_{\ceci-1}\in X'.
\end{aligned}
\end{gather}

Finally, the equation (\ref{aisaprime}) means that the local maxima of $X$ are identical to those of $X'$, and then the property (1-2) shows that $X=X'$, the statement (1-3).

\subsection{Proof of Step 2}
\label{s:6.2}

\subsubsection*{Difference of the characterizing numbers}
Under the condition (\ref{Xpartition}), we consider the difference of
characterizing numbers $N(a_{q}+\mu(s))$ of two boxes both located on 
the rightmost edges of Young diagrams. The number for either box has of the form
\ba
x+(\lambda_{\vec Y}(x)-1)(N+M),
\ea
and their difference becomes
\ba\label{diffxy}
(x-y)+(\lambda_{\vec Y}(x)-\lambda_{\vec Y}(y))(N+M).
\ea
It does not vanish if $x<_{X} y$ or $y<_{X} x$ by the definition of $<_{X}$. 
Suppose that the factor (\ref{diffxy}) vanishes, it implies
\ba\label{diffxy:1}
y=x+(\lambda_{\vec Y}(x)-\lambda_{\vec Y}(y))N+(\lambda_{\vec Y}(x)-\lambda_{\vec Y}(y))M\,,
\ea
which leads to either $x\leq_X y$ or $x\geq_Xy$. We can see the only consistent solution is $x=y$. 
This result implies the factors $a_p-a_q+A_k(Y_p)-A_\ell(Y_q)$ and 
$a_p-a_q+B_k(Y_p)-B_\ell(Y_q)$ in the denominator of $\Lambda^{\pm}$, (\ref{Lambda}), 
vanish if and only if for $(p,k)=(q,l)$.

Similarly, if the factor
\ba\label{diffxy:2}
(x-y)+(\lambda_{\vec Y}(x)-\lambda_{\vec Y}(y))(N+M)+(N+M)=(x-y)+(\lambda_{\vec Y}(x)-\lambda_{\vec Y}(y)+1)(N+M)
\ea
vanishes, we can only have $x<_{X}y$ or $x>_{X}y$, and then the condition (\ref{Xpartition}) gives $x>_{X}y$ and $\lambda_{\vec Y}(x)=\lambda_{\vec Y}(y)$. As a result, if (\ref{diffxy:2}) vanishes, we have $y=x+N+M$ and $\lambda_{\vec Y}(x)=\lambda_{\vec Y}(y)$. 
In particular, we have $\lambda_{\vec Y}(x)=\lambda_{\vec Y}(x+N)$. 
This, in turn, implies $a_p-a_q+A_k(Y_p)-B_\ell(Y_q)$ in the product (\ref{Lambda}) does not vanish since the factor $a_{p}+A_{k}(Y_{p})$ comes from a row $x$ with $\lambda_{\vec{Y}}(x)<\lambda_{\vec{Y}}(x+N)$. 

On the other hand, the factor $a_p-a_q+A_k(Y_p)-B_\ell(Y_q)+\xi$ in (\ref{Lambda}) corresponds to
\ba\label{diffxy:3}
(x-y)+(\lambda_{\vec Y}(x)-\lambda_{\vec Y}(y))(N+M)+N\,.
\ea
Through the same discussion, the only possibility for this factor to vanish reduces to 
$y=x+N$ and $\lambda_{\vec{Y}}(x)=\lambda_{\vec{Y}}(x+N)$. Therefore $a_p-a_q+A_k(Y_p)-B_\ell(Y_q)+\xi$ never vanishes.

As a result, the $\Lambda^{+}$ (resp. $\Lambda^{-}$)  in (\ref{Lambda}) vanishes if and only if an upper-right factor $a_p-a_q+A_k(Y_p)-A_\ell(Y_q)+\xi$ (resp. $a_p-a_q+B_k(Y_p)-B_\ell(Y_q)+\xi$) vanishes. Note that these factors correspond to 
\ba\label{diffxy:4}
(x-y)+(\lambda_{\vec Y}(x)-\lambda_{\vec Y}(y))(N+M)+M\,.
\ea
which vanishes if $y=x+M$ and $\lambda_{\vec{Y}}(x)=\lambda_{\vec{Y}}(x+M)$, which can be shown by the same discussion.

\subsubsection*{Proof of the statement (3)}
We show that no vector of Young diagrams which breaks the condition (\ref{Xpartition}) appears in the highest weight representation.
This means that,  for the finiteness of $\Lambda^{\pm}$, the highest weight representation is well-defined.  Also, it shows that the module is spanned by {\it some of} the integer partitions satisfying (\ref{Xpartition}).
 
Let $\vec Y$ be an $N$-tuple Young diagrams whose $\lambda_{\vec{Y}}$ satisfies (\ref{Xpartition}). Then let us denote $\vec{Y}_{+}$ be another $N$-tuple obtained by adding one box to $\vec{Y}$ at the position $a_{p}+A_{k}(Y_{p})$. 
Assume that $\lambda_{\vec{Y}_{+}}$ does not satisfy the condition (\ref{Xpartition}). Then we should have $\lambda_{\vec Y}(x+M)=\lambda_{\vec Y}(x)$ for $x=N(a_{p}+\mu(s^{(p)}_{k}+1,1))$ such that $x+M\in X$, 
where 
$Y_p=(r^{(p)}_t, s^{(p)}_t)_{t=1}^{f_p}$ is the rectangle decomposition of $Y_{p}$.
Since the condition (\ref{Xpartition}) holds for the other rows, we see that in the row with $x+M$ we can add a box to obtain another $N$-tuple Young diagrams. 
In other words, it is a hollow place of the Young diagram belonging to the row with $x+M$.
This means that there exist $q\neq p$ and $l$ such that $N(a_{q}+A_{l}(Y_{q}))=N(a_{p}+A_{k}(Y_{p}))+M$.
This equation forces a factor in the numerator of $\Lambda^{(k,+)}_{p}(\vec a, \vec{Y})$ to vanish, which prevents
 $\ket{\vec a, \vec{Y}_{+}}$ to appear in the representation.
 Therefore, the action of $D_{1,l}$ closes in the subspace spanned by such desired vectors. So is $D_{-1, l}$, shown by a similar discussion.
 
We complete the proof of the statement (3). What remains is that  {\it all} the states labeled by integer partitions satisfying (\ref{Xpartition}) appear in the highest weight representation. However, it is immediate as follows. With the notations above, assume that $\lambda_{\vec{Y}_{+}}$ also satisfies (\ref{Xpartition}), and then $\Lambda^{(k,\pm)}_{q}(\vec a, \vec{Y})$ does not vanish for $\lambda_{\vec Y}(x+M)>\lambda_{\vec Y}(x)$ and a discussion around (\ref{diffxy:4}).
Since no two arbitrary $a_{q}+A_l(Y_q)$ are identical, then an appropriate linear combination of $D_{1,l}$ maps $\ket{\vec{a}, \vec{Y}}$ to $\ket{\vec{a}, \vec{Y}_{+}}$.

\subsubsection*{Proof of the statement (2)}
\begin{figure}
\centering
\begin{tikzpicture}[scale=.4]
\coordinate (A)at(0,0);
\coordinate (B)at(0,-10);
\coordinate (C)at(5,-10);
\coordinate (D)at(5,-7);
\coordinate (E)at(10,-7);
\coordinate (F)at(10,-3);
\coordinate (G)at(15,-3);
\coordinate (H)at(15,0);
\node[] at (-7,-5) {$\Lambda^{(k,+)}_{p}(\vec a, \vec{Y})^2=\beta\times\prod_{x}$};
\node[right] at (7,-10) {, where \ytableausetup{centertableaux,boxsize=1em}\begin{ytableau} \pm \end{ytableau}\ytableausetup{centertableaux,boxsize=.6em} means $\left(N(a_{q}-a_{p}+\mu(\ydiagram{1})-A_k(Y_{p}))\right)^{\pm1}.$};
\draw (A)--(B)--(C)--(D)--(E)--(F)--(G)--(H)--(A);
\draw[] (0,-4) rectangle (10,-5);
\draw[] (9,-3) rectangle node{$+$} (10,-4);
\draw[] (9,-4) rectangle node{$-$} (10,-5);
\draw[] (10,-4) rectangle node{$-$} (11,-5);
\draw[] (10,-5) rectangle node{$+$} (11,-6);
\draw[->,bend right,ultra thick, in=-90, out=-90] ([yshift=2.5cm]1,-1)node[right]{A row labeled by $x\neq N(a_{p}+\mu(s_{k+1}^{(p)},1))$}to([xshift=.5cm,yshift=-.5cm]0,-4);
\end{tikzpicture}
\caption{Diagrammatic expression of $\Lambda^{(k,+)}_{p}(\vec a, \vec{Y})$}\label{tetris}
\end{figure}
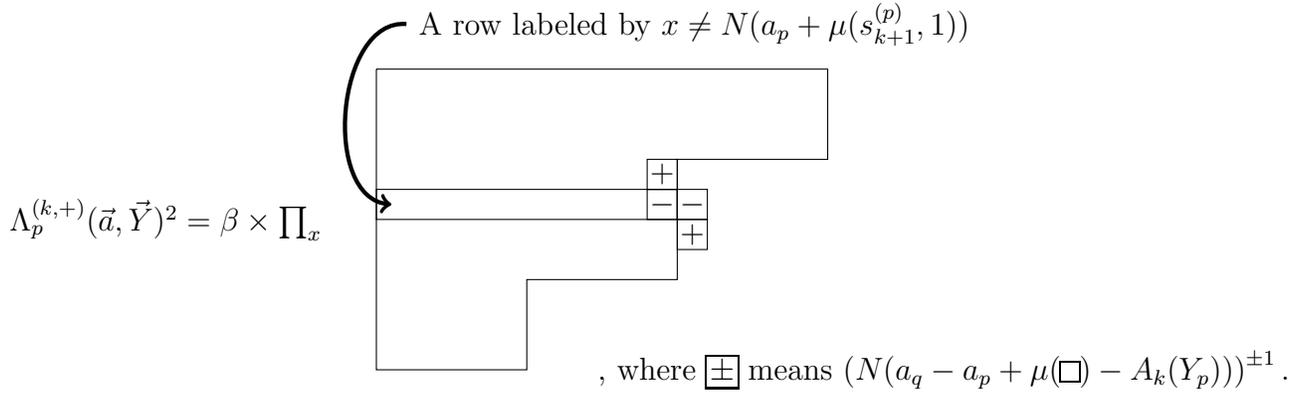

Finally we would like to establish that the action of $D_{\pm 1,l}$ is consistent with the shuffling of rows.
We note that we can rewrite the products appeared in the definition (\ref{Lambda}) of $\Lambda^{\pm}$ to some products over all the rows but one
\footnote{The missing one comes from the upper right numberator factor in $\Lambda^\pm$, which corresponds to $Y_p=Y_q$ and $k=\ell$. This missing factor is merely a constant $\left((\pm)M\right)^{-1/2}$, which reduces to the overall rescaling factor 
$(M/N)^{-1/2}$}\footnote{To avoid discussions about zeroth or poles, one may shift all the factors in the tetrimino by a small parameter $h$. After the product is done, we take a limit $h\to0$ and obtain the original $\Lambda^{\pm}$.} (see Figure \ref{tetris}).
This implies that $D_{\pm 1,l}$ and the shuffling commute up to an overall rescaling (\ref{rescale2}) of $D_{\pm1,l}$ with $\gamma=1$ under the level-rank duality. 

In summary, it has been shown that one can obtain the irreducible representation from the level-0 state just by setting parameters of the SH$^c$ representation to fit minimal models with the level-rank duality. Then the level-rank duality is realized as an intertwining map which shuffles rows of Young diagrams, following the characterizing set $X$.

%%%%%%%%%%%%%%%%%%%%%%%%%%%%%%%%%
\section{Poset and partition function}
\label{section6a}
For the minimal models with the level-rank duality, we have understood that 
the excited states in the SH$^c$ module  are described by
the partitions $\lambda: X\to \mathbb{N}$.
Recall we have introduced a partial order $<_{X}$ to the set $X$ by
\ba\label{Xordering}
x\ge_{X}y \Leftrightarrow {}^{\exists} i_{N}, i_{M}\in\mathbb{N},\ x=y-Ni_{N}-Mi_{M},
\ea
and the partition $\lambda$ is defined by
\ba\label{Ppartition}
x\ge_{X}y \Rightarrow \lambda(x)\le\lambda(y).
\ea
Denoting the set of all such partitions by $\mathcal{A}(X)$, now we consider the 
Hilbert series $Z_{X}(q)$ of the representation graded by its level, which is defined by
\ba
Z_{X}(q)=\sum_{\lambda\in\mathcal{A}(X)}q^{|\lambda|}\,.
\ea
We note that it defines a new way to count the number of states
for (the special cases of) the minimal models.

The character of the general minimal model was derived
through the coset construction \cite{frenkel1992characters}
and N-Burge condition \cite{Feigin:2010qea}.
For our special cases, it is given in the infinite product form \cite{Altschuler:1990th}.
While it is well-known, the new counting method may provide
a new way to write the character which produces nontrivial identities.

For the set  $X=\mathbb{N}$ with the canonical order, 
the computation of the partition function is elementary:
\begin{gather}\label{HilbCountYoung}
Z_{\mathbb{N}}(q)=\prod_{i=1}^{\infty}\sum_{n_{i}=0}^{\infty}q^{in_{i}}=\frac{1}{(q; q)_{\infty}},
\end{gather}
where $(a;q)_j=\prod_{k=0}^{j-1}(1-aq^k)$ is the $q$-Pochhammer symbol. 

How do we calculate partition function for $Z_{X}(q)$ for the partially ordered set?
In mathematics, the integer partitions on the partially ordered set are well studied and called {\it P-partitions}. 
Before we introduce some results in the book \cite{Stanley:1986:EC:21786}, 
we explain how to organize $\mathcal{A}(X)$ to calculate the Hilbert series by dealing with a toy example.

\subsection{A toy example}
\begin{figure}
\centering
\begin{tikzpicture}[scale=.4]
\coordinate (p1)at(0,0);
\coordinate (p2)at(-5,-5);
\coordinate (p3)at(5,-5);
\foreach \p in {1,2,3}
\draw[fill] (p\p) circle[radius=.2] node[left] {$p_{\p}$};
\foreach \p in {2,3}
\draw (p1) -- (p\p);
\end{tikzpicture}
\caption{The Hasse diagram of $P=\{p_{1}, p_{2}, p_{3}\}$}\label{Hasse:1}
\end{figure}
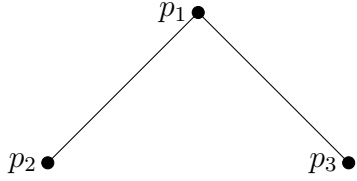

Let us consider a partially ordered set $P=\{p_{1}, p_{2}, p_{3}\}$ of three points as an example, whose ordering is given by $p_{1}<_{P}p_{2}$ and $p_{1}<_{P}p_{3}$. Its Hasse diagram\footnote{Given a partially ordered set $P$, its Hasse diagram is defined by assigning a point for each $x\in P$ and drawing a line between two points representing $x, y\in P$ if $x<_{P}y$ and there is no element $z\in P$ such that $x<_{P}z<_{P}y$. Clearly, if $x<_{P}y$, $x$ and $y$ lies in the same connected component of the graph.} is represented as Figure \ref{Hasse:1}.
We denote by $\mathcal{A}(P)$ the set of partitions $\lambda: P\to\mathbb{N}$ satisfying $|\lambda|<\infty$ and (\ref{Ppartition}) for $X=P$, and call its element a P-partition. For example, $(p_1, p_2, p_3)\mapsto (2,1,0), (2,0,1)$ are P-partitions.

\begin{figure}
\centering
\begin{tikzpicture}[scale=1]
\coordinate (p1)at(0, -.5);
\coordinate (p3)at(0, -1.5);
\coordinate (p2)at(0, -2.5);
\foreach \p in {1,2,3}
\node[left] at (p\p) {$p_{\p}$};
\foreach \p in {1,2}
\node[left] at (-.1,-\p) {\rotatebox{-90}{$<_{\pi}$}};
\draw (0,0)--(3,0);
\draw (0,-1)--(3,-1);
\draw (0,-2)--(2,-2);
\draw (0,-3)--(1,-3);
\draw (0,0)--(0,-3);
\draw (1,0)--(1,-3);
\draw (2,0)--(2,-2);
\draw (3,0)--(3,-1);
\end{tikzpicture}
\caption{The P-partition $(3,1,2)$ expressed by a Young diagram with a total ordering $<_\pi$}\label{PtoYoung}
\end{figure}
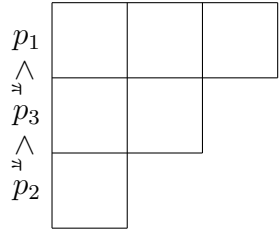

Note that, for each P-partition $\lambda$, there is a total ordering $<_{\pi}$ of $P$ satisfying
\ba
x<_{P}y\Rightarrow x<_{\pi}y
\ea
and
\ba
x<_{\pi}y\Rightarrow \lambda(x)\ge\lambda(y).
\ea
In other words, for given a P-partition, we can arrange all the elements of $P$ into a line compatible with the ordering $<_{P}$ and then the partition is expressed by a Young diagram whose rows represent the linearized $P$ and $\lambda(x)$ is equal to the number of boxes in the row with $x$ for each $x\in P$.
For example, for the P-partition $(p_1, p_2, p_3)\mapsto (3,1,2)$, we linearize $P$ by a total order $p_{1}<_{\pi}p_3<_{\pi}p_2$ and then the corresponding diagram is expressed by a Young diagram appeared in Figure \ref{PtoYoung}.

Then, in order to compute the Hilbert series of $P$, we want to correspond a 
P-partition to a pair of a total ordering compatible to $<_{P}$ and a Young diagram. 
We have to be careful to avoid overcountings. For example, for a P-partition 
$(p_1, p_2, p_3)\mapsto (2,1,1)$, there are two total ordering $p_{1}<p_{2}<p_{3}$ and 
$p_{1}<p_{3}<p_{2}$ both of which give the same Young diagram \ydiagram{2,1,1}. 
To count without overlaps, we fix one total ordering, say $p_{1}<p_{2}<p_{3}$, 
as the canonical one, and correspond that partition to the pair of this 
canonical ordering and the Young diagram \ydiagram{2,1,1}. 
As a result, we separate $\mathcal{A}(P)$ into the two set $S_{1}, S_{2}$:
\begin{gather}
\begin{aligned}
\mathcal{A}(P)&=S_{1}\sqcup S_{2},\\
S_{1}&\equiv \{\lambda\in\mathcal{A}(P)\ |\ \lambda(p_{1})\ge\lambda(p_2)\ge\lambda(p_3)\},\\
S_{2}&\equiv \{\lambda\in\mathcal{A}(P)\ |\ \lambda(p_{1})\ge\lambda(p_3)>\lambda(p_2)\}.
\end{aligned}
\end{gather}

\begin{figure}
\centering
\begin{tikzpicture}[scale=.6]
\coordinate (p1)at(0, -.5);
\coordinate (p3)at(0, -1.5);
\coordinate (p2)at(0, -2.5);
\foreach \p in {1,2,3}
\node[left] at (p\p) {$p_{\p}$};
\draw (0,0)--(0,-3)--(2,-3)--(2,-2)--(4,-2)--(4,-1)--(6,-1)--(6,0)--(0,0);
\draw[bend right] (2,-2) to node[below]{$>0$} (4,-2);
\node at (7,-1.5) {$\Rightarrow$};
\draw[xshift=8cm] (0,-2)--(0,-3)--(2,-3)--(2,-2)--(4,-2)--(4,-1)--(6,-1)--(6,0)--(1,0)--(1,-2)--cycle;
\draw[dashed](9,-2)--(10,-2);
\draw[bend right, xshift=8cm] (2,-2) to node[below]{$>0$} (4,-2);
\draw[xshift=-.2cm, yshift=.2cm] (8,0) rectangle (9,-1);
\draw[xshift=-.2cm, yshift=.2cm] (8,-1) rectangle (9,-2);
\node at (11, -1) {$\leftarrow$};
\node at (15,-1.5) {$\Rightarrow$};
\draw[xshift=16cm] (0,0)--(0,-3)--(2,-3)--(2,-2)--(3,-2)--(3,-1)--(5,-1)--(5,0)--(0,0);
\draw[bend right, xshift=16cm] (2,-2) to node[below right]{$\ge0$} (3,-2);
\node at (22,-1.5) {$+$};
\draw[xshift=15cm, yshift=-.5cm] (8,0) rectangle (9,-1);
\draw[xshift=15cm, yshift=-.5cm] (8,-1) rectangle (9,-2);
\end{tikzpicture}
\caption{Modification of the condition $\lambda(p_3)>\lambda(p_2)$ in $S_{2}$}\label{PtoYoung:2}
\end{figure}
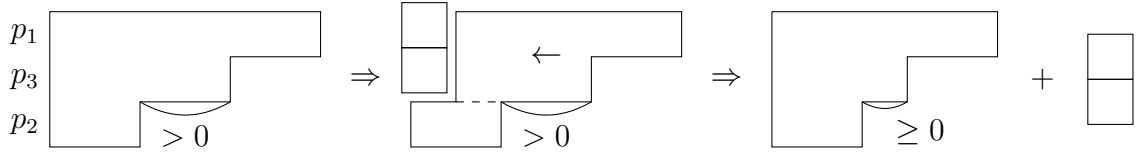

Now we can compute the Hilbert series through a similar way we performed in (\ref{HilbCountYoung}) for the case with $X=\mathbb{N}$. 
On the $S_{1}$ side, we have
\ba
\sum_{\lambda\in S_{1}}q^{|\lambda|}=\frac{1}{(q;q)_{3}}.
\ea
On the other side, through a diagrammatic consideration in Figure \ref{PtoYoung:2} for the condition $\lambda(p_3)>\lambda(p_2)$ in the definition of $S_{2}$,
we have
\ba\label{S2computation}
\sum_{\lambda\in S_{2}}q^{|\lambda|}=q^{2}\sum_{\lambda\in S_{1}}q^{|\lambda|}=\frac{q^2}{(q;q)_{3}}.
\ea
Then we have $Z_{P}(q)=(1+q^{2})/(q;q)_{3}$. Note that the $q^{2}$ appeared in the numerator of (\ref{S2computation}) comes from the fact that the $p_{3}$, the $2^{\text{nd}}$ lowest element in $P$ with respect to the ordering $p_{1}<p_{3}<p_{2}$, is larger than the next element $p_{2}$ with respect to the fixed canonical ordering $p_{1}<p_{2}<p_{3}$. In other words, this factor represents where a total ordering of $P$ differs from the canonical total ordering.

\subsection{General cases}
This example gives us a lesson on how to compute the Hilbert series for a general partially ordered set 
which is bounded below with finite local minima and whose Hasse diagram is connected
\footnote{If the Hasse diagram is not connected, the Hilbert series is given by the product of the series for each connected component of the partially ordered set.}:
\begin{enumerate}
\item
Define $\mathcal{L}(X)$ as the set of all the total orderings compatible to the ordering $<_{X}$, known for mathematicians as the {\it Jordan-H\"{o}lder} set of $X$. 
\item
Fix an element of $c\in\mathcal{L}(X)$ as the canonical ordering, and then for each $\pi\in\mathcal{L}(X)$ define
\ba
S_{\pi}=\left\{\lambda\in\mathcal{A}(X) \middle|\ \begin{aligned}x<_{\pi}y &\Rightarrow \lambda(x)\ge\lambda(y)\\   x>_{c}x+1 &\Rightarrow \lambda(x)>\lambda(x+1)\end{aligned}\right\},
\ea
where $x+1$ is the next element to $x$ with respect to the ordering $\pi$.
\item
Denoting the set of all the positions where two ordering $\pi\in\mathcal{L}(X)$ and $c$ differ by
\ba
D_{\pi}=\{\ j\in\mathbb{N}^{+}|\ x_{j}>_{c}x_{j+1}\},
\ea
where $x_{j}$ represents the $j$-th lowest element of $X$ with respect to $\pi$,  we have\footnote{The corresponding statement in the book \cite{Stanley:1986:EC:21786} is for partially ordered sets with {\it finite} cardinality. However, it can be enlarged for our infinite cases since boxes are piled up from the bottom.} \cite{Stanley:1986:EC:21786}
\begin{gather}
\mathcal{A}(X)=\bigsqcup_{\pi\in\mathcal{L}(X)}S_{\pi},\\
Z_{X}(q)=\sum_{\lambda\in\mathcal{A}(X)}q^{|\lambda|}=\frac{\sum_{\pi\in\mathcal{L}(X)}\prod_{j\in D_{\pi}}q^{j}}{(q;q)_{|X|}},\label{ZX}
\end{gather}
where $|X|$ is the cardinality of $X$.
\end{enumerate}

Here we rewrite $D_{\pi}$ as follows. For a given $\pi\in\mathcal{L}(X)$, we denote by $\pi(i)=j$ that the $i$-th lowest element with respect to $\pi$ is the $j$-th lowest one with respect to the canonical ordering. Then we have
\begin{gather}
D_{\pi}=\{\ i\ |\ \pi(i)>\pi(i+1)\}.
\end{gather}

\subsection{Application to minimal model: $(2,3)$ case}
We have seen that the computation of the partition function for the poset 
is reduced to find the Jordan-H\"{o}lder set of $X$.
It is, however, difficult to find it for the poset $X$ for minimal models at this moment.
Fortunately, for the simplest example $(N,M)=(2,3)$, we can explicitly find them
and compute $Z_{X}(q)$ explicitly.  

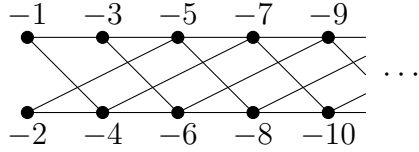
\begin{figure}
\centering
\begin{tikzpicture}[scale=1]
\foreach \q in {0,1}
\foreach \p in {0,1,...,4}
\pgfmathtruncatemacro{\i}{2*\p+\q+1}
\coordinate (\i)at (\p,-\q);
\foreach \r in {1,3,...,9}
\draw[fill] (\r) circle[radius=.08] node[above] {$-\r$};
\foreach \r in {2,4,...,10}
\draw[fill] (\r) circle[radius=.08] node[below] {$-\r$};
\foreach \s in {1,2,...,7}
\pgfmathtruncatemacro{\j}{\s+3}
\draw (\s)--(\j);
\foreach \s in {1,2,...,8}
\pgfmathtruncatemacro{\j}{\s+2}
\draw (\s)--(\j);
\draw (8)--([xshift=1.5cm,yshift=.75cm]8);
\draw (9)--([xshift=.5cm]9);
\draw (10)--([xshift=.5cm]10);
\draw (9)--([xshift=.5cm,yshift=-.5cm]9);
\draw (10)--([xshift=.5cm,yshift=.25cm]10);
\node at (5, -.5) {$\cdots$};
\end{tikzpicture}
\caption{The Hasse diagram of $X$ for $\Delta=-\frac{1}{5}$}
\label{Hasse:2}
\end{figure}

For the $\Delta=-1/5$ case ($\tn=(1,2)$), we have $\{N a_q +M\}= \{-1, -2\}$ on the $N=2$ side. 
Then we obtain Figure \ref{Hasse:2} as the Hasse diagram of the corresponding ordered set $X$.
Now we fix as the canonical ordering a total ordering where the $i$-th lowest element is $-i\in X$. 
Then since $\pi$ is an element of $\mathcal{L}(X)$, we have
\ba
\pi(1)=1,\quad\text{or}\quad(\pi(1), \pi(2))=(2,1)\,.
\ea
In either case, the lowest element or pair with respect to $\pi$ is selected as the lowest one with respect to the canonical ordering.
Note that the subset obtained by the subtraction of a set $\{ x\in X| x<_{c}-j\}$ from $X$ for a given integer $j$ gives the same Hasse diagram\footnote{This is why we can determine $\mathcal{L}(X)$ for the Lee-Yang edge singularity.} as the one with the original $X$.
Therefore a similar analysis goes well for more higher elements of $X$, and then we can identify $\mathcal{L}(X)$ with a set of integers
\ba
\{1\le t_{1}<t_{2}<\dots<t_{k}\ | k\in\mathbb{N}, t_{i}-t_{i-1}\ge2\}, 
\ea
where the isomorphism is given by $\pi(t_{i})>\pi(t_{i}+1)$. Surprisingly, the above set is none other than the one of integer partitions with the Rogers-Ramanujan identities (see \cite{gordon1961combinatorial, BA06553742})!  As a result, we have
\begin{gather}\label{partition-1/5}
\begin{aligned}
Z_{X}^{(\Delta=-\frac{1}{5})}(q)&=\frac{1}{(q;q)_{\infty}}\sum_{k=0}^{\infty}\sum_{\substack{1\le t_{1}<\dots<t_{k}\\ t_{i}-t_{i-1}\ge2}}q^{\sum_{j=1}^{k}t_{j}}\\
&=\frac{1}{(q;q)_{\infty}}\sum_{k=0}^{\infty}\sum_{0\le t_{1}\le\dots\le t_{k}}q^{k+\sum_{j=0}^{k-1}2j+\sum_{j=1}^{k}t_{j}}\\
&=\frac{1}{(q;q)_{\infty}}\sum_{k=0}^{\infty}q^{k^{2}}\sum_{0\le t_{1}\le\dots\le t_{k}}q^{\sum_{j=1}^{k}t_{j}}\\
&=\frac{1}{(q;q)_{\infty}}\sum_{k=0}^{\infty}\frac{q^{k^{2}}}{(q;q)_{k}}.
\end{aligned}
\end{gather}
The partition function for this case is known:
\ba
Z_{X}^{(\Delta=-\frac{1}{5})}(q)&=&\frac{1}{(q;q)_{\infty}(q;q^5)_{\infty}(q^4;q^5)_{\infty}}=\frac{(q^2;q^5)_{\infty}(q^3;q^5)_{\infty}(q^5;q^5)_{\infty}}{(q;q)^2_{\infty}}\,.
\ea
The partition function obtained by two methods gives rather different form.
It is, however, the same due to
the first Rogers-Ramanujan identity \cite{ramanujan1919proof}.
In a sense, the computation of partition through the poset suggests 
Rogers-Ramanujan identity.

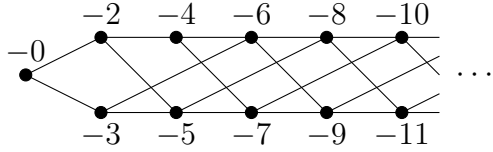
\begin{figure}
\centering
\begin{tikzpicture}[scale=1]
\foreach \q in {0,1}
\foreach \p in {0,1,...,4}
\pgfmathtruncatemacro{\i}{2*\p+\q+2}
\coordinate (\i)at (\p,-\q);
\coordinate (0)at (-1,-.5);
\foreach \r in {3,5,...,11}
\draw[fill] (\r) circle[radius=.08] node[below] {$-\r$};
\foreach \r in {0,2,...,10}
\draw[fill] (\r) circle[radius=.08] node[above] {$-\r$};
\foreach \s in {0,2,3,...,8}
\pgfmathtruncatemacro{\j}{\s+3}
\draw (\s)--(\j);
\foreach \s in {0,2,3,...,9}
\pgfmathtruncatemacro{\j}{\s+2}
\draw (\s)--(\j);
\draw (10)--([xshift=.5cm]10);
\draw (11)--([xshift=.5cm]11);
\draw (10)--([xshift=.5cm,yshift=-.5cm]10);
\draw (11)--([xshift=.5cm,yshift=.25cm]11);
\draw (9)--([xshift=1.5cm,yshift=.75cm]9);
\node at (5, -.5) {$\cdots$};
\end{tikzpicture}
\caption{The Hasse diagram for $\Delta=0$}
\label{Hasse:3}
\end{figure}

We study the other case $\Delta=0\ (\tn=(0,3), Na_{q}+M=0,-3)$. The Hasse diagram of the corresponding partially ordered set is given in Figure \ref{Hasse:3}.
We take the total ordering $0<_{c}-2<_{c}-3<_{c}\dots$ as the canonical one and then we find $\pi(1)=1$ for any $\pi\in\mathcal{L}(X)$. Note that we see from Figure \ref{Hasse:3} that the Hasse diagram of the subset $X\setminus\{0\}$ is identical to the Figure \ref{Hasse:2}. Therefore we can apply the previous discussions to higher elements of $X$. As a result, we identify $\mathcal{L}(X)$ with
\ba
\{1<t_{1}<t_{2}<\dots<t_{k}\ | k\in\mathbb{N}, t_{i}-t_{i-1}\ge2\}.
\ea
Therefore, we have
\begin{gather}\label{partition-0}
\begin{aligned}
Z_{X}^{(\Delta=0)}(q)&=\frac{1}{(q;q)_{\infty}}\sum_{k=0}^{\infty}\sum_{\substack{1< t_{1}<\dots<t_{k}\\ t_{i}-t_{i-1}\ge2}}q^{\sum_{j=1}^{k}t_{j}}\\
&=\frac{1}{(q;q)_{\infty}}\sum_{k=0}^{\infty}\sum_{0\le t_{1}\le\dots\le t_{k}}q^{2k+\sum_{j=0}^{k-1}2j+\sum_{j=1}^{k}t_{j}}\\
&=\frac{1}{(q;q)_{\infty}}\sum_{k=0}^{\infty}\frac{q^{k^{2}+k}}{(q;q)_{k}}\\
&=\frac{1}{(q;q)_{\infty}(q^2;q^5)_{\infty}(q^3;q^5)_{\infty}}=\frac{(q;q^5)_{\infty}(q^4;q^5)_{\infty}(q^5;q^5)_{\infty}}{(q;q)^2_{\infty}}.
\end{aligned}
\end{gather}
We have met the second Rogers-Ramanujan identity \cite{ramanujan1919proof}
when we go to the last line.

We should comment on these results. (1) It was already known in \cite{Kac-Wakimoto} that the Lee-Yang case relates to the Rogers-Ramanujan identities. 
We only revisit the same result from another path, the P-partition counting. 
(2) Note that the two Hilbert series both include the extra factor $(q;q)_\infty$ in the denominators in (\ref{partition-1/5}), 
(\ref{partition-0}), compared to the results in \cite{Kac-Wakimoto}. This factor seems to come from the fact that SH$^c$ contains not only the $\cW_{N}$-algebra but also the Heisenberg (or $U(1)$ current) algebra.

\subsection{Conjectures from the general $(N,M)$ cases}
As noted before, we cannot obtain the Jordan-H\"older set for the poset $X$
for the general cases which makes the computation of P-partition difficult.
Since we have established that the partition for the poset $X$ 
is equivalent to the Hilbert space of minimal models relevant in the level-rank duality,
the partition function for the poset should be equal to the 
character formula for minimal models \cite{Kac-Wakimoto}:
\ba\label{ZvsChi}
Z_{X}(q)=\frac{\chi_{N}(q)}{(q;q)_{\infty}(q^{N+M};q^{N+M})_{\infty}}\,,
\ea
where, for $\cW_N$ minimal models with the level-rank duality, the character $\chi_{N}(q)$ takes of the form \cite{Altschuler:1990th}
\ba
\chi_N(q)=\frac{(q^{N+M}; q^{N+M})^{N}_{\infty}\prod_{i<j}(q^{x_i-x_j};q^{N+M})_{\infty}\prod_{i>j}(q^{N+M+x_i-x_j};q^{N+M})_{\infty}}
{(q;q)^{N-1}_{\infty}}\,,\label{chi-for}
\ea
where $x_i=\sum_{j=i}^N n_j$.
It is known in \cite{Altschuler:1990th} that the $\chi_{N}(q)$ is invariant under replacements $N\mapsto M$ and $x\mapsto y$ (see subsection \ref{LRdual}) corresponding to the level-rank duality, which gives the character on the $\cW_M$-algebra side.

In summary, we have the following conjecture:
\begin{conjecture}
On the Hilbert series of the SH$^c$ module with the level-rank duality, the following identiy holds;
\ba\label{ZvsChi-f}
\sum_{\lambda\in\mathcal{A}(X)}q^{|\lambda|}=\frac{(q^{N+M}; q^{N+M})^{N-1}_{\infty}\prod_{i<j}(q^{x_i-x_j};q^{N+M})_{\infty}\prod_{i>j}(q^{N+M+x_i-x_j};q^{N+M})_{\infty}}
{(q;q)_{\infty}^{N}}\,,
\ea
where the partially ordered set $X$ is defined by (\ref{defX}) and (\ref{Xordering}), and the $\mathcal{A}(X)$ is the set of all the P-partitions (\ref{Ppartition}) over $X$.
\bsq
\end{conjecture}
\section{Constraints on the Hilbert space of general minimal models: $N$-Burge condition}\label{section7}

In this section, we first derive the $N$-Burge condition, 
which was obtained in \cite{burge1993restricted, bershtein2014agt, alkalaev2014conformal} 
for the $N=2$ case and in \cite{Belavin:2015ria} from a family of zeros in $\Lambda^+$. 
We note that similar result was obtained a few years ago in \cite{Feigin:2010qea} for the q-deformed case.
Then we show that this is the sufficient restriction to get the whole submodule 
without null states in the representation of SH$^c$ corresponding to a general minimal model. 
As the proof is almost parallel to those in section \ref{section6}, 
our explanation will be focused on the points where the generalization is nontrivial.

\subsection{$N$-Burge condition}
The $N$-Burge condition in our convention reads
\ba
Y_{i,\mathrm{R}}-Y_{i+1,\mathrm{R}+(n_i -1)} \geq -(n'_i -1).\label{N-Burge}
\ea
Here, $Y_{i,\mathrm{R}}$ denotes the number of boxes in the $\mathrm{R}$-th row of the $i$-th Young diagram.

In order to derive it from the SH$^c$, we consider the factors in 
\ba
\Lambda^{(k,+)}_p(\vec a,\vec Y) &=& \left(
\prod_{q=1}^N \left(\prod_{\ell=1}^{f_q} \frac{
a_p-a_q+A_k(Y_p)-B_\ell(Y_q)+\xi
}{
a_p-a_q+A_k(Y_p)- B_\ell(Y_q)
}{\prod}_{\ell=1}^{\prime f_q +1}\frac{a_p-a_q+A_k(Y_p)- A_\ell(Y_q) -\xi}{
a_p-a_q+A_k(Y_p)-A_\ell(Y_q)
}
\right)\right)^{1/2}\,.\nn
\ea
We notice that the condition for the numerator in the second product to vanish is $a_p-a_q+A_k(Y_p)- A_\ell(Y_q) -\xi = 0$ for some $p$, $q$, $k$ and $\ell$. 
Here, we focus on the case $p = q+1$.
\ba
0 &=& a_{q+1}-a_{q}+A_k(Y_{q+1})- A_\ell(Y_{q}) -\xi\nn\\
&=& \beta(r^{q+1}_{k-1} -  r^{q}_{l-1} -n'_q +1 ) + (s^{q}_{l} - s^{q+1}_k +n_q -1)
\ea
To make $r^{q+1}_{k-1}=Y_{q+1,R}$ for some $R$, we have to set $s^{q+1}_k=R+1$. 
Then considering the situation with $s^q_l=R'-1=R-n_q$, we have 
\ba
Y_{q,\mathrm{R}'}-Y_{q+1,\mathrm{R}'+(n_q -1)} = -(n'_q -1).
\label{aafactor}
\ea
The highest weight vector with all empty Young diagrams implies the inequality in (\ref{N-Burge}), we thus know (\ref{N-Burge}) is the right condition to impose 
to extract the submodule including the highest weight. 

When we go back to the special case of the level-rank duality, the right-hand side of (\ref{N-Burge}) vanishes. Those two rows under consideration 
have respectively the characterizing integers $Na_i+M-(R-1)N$ and $Na_{i+1}+M-(R+\tilde{n}_i-1)N$. Their difference is exactly $M$, that is to say, 
(\ref{N-Burge}) reduces to 
\ba
\lambda(x+M)\geq\lambda(x)\,.\nn
\ea
We thus see that in this specific case, as we showed before, 
the $N$-Burge condition together with the requirement that all sates are labeled by Young diagrams, 
i.e. $\lambda(x+N)\geq \lambda(x)$, determine the spectrum completely. We also expect that a similar discussion to that in section 6 applies to show the $N$-Burge condition plus 
the requirement from the Young diagram filter out all the states generated from the highest weight with $D_{1,l}$. 

\subsection{Sufficiency of the $N$-Burge condition in SH$^c$}
Let us again introduce the characterizing number $x=pa_i+q-ps\in X$ for the $s$-th row in the $i$-th Young diagram. 
Then the $N$-Burge condition and the Young diagram requirement can be rewritten to
\begin{gather}\label{two-conditions}
\begin{aligned}
\lambda(x-p)&\leq\lambda(x)\,,\\
\lambda(x-q\tilde{n}'_i-(q-p))&\leq\lambda(x)+\tilde{n}'_i\,.
\end{aligned}
\end{gather}
We define the ordering in $X$ as follows: if there exist positive integers $l$, and $^\exists i<^\exists j$, s.t. 
\ba
y=x-lp-(j-i)(q-p)-q\sum_{k=i}^{j-1}\tilde{n}'_k\,,
\ea
then $x<_X y$. Under this ordering, we can reproduce almost the whole part of the proof we provided in section 
\ref{s:6.2}.
However, in (\ref{diffxy:1}), we used the fact that the coefficients before $N$ and $M$ when we connect $x$ and $y$ are 
integers. In more general cases, $\lambda(x)-\lambda(y)$ should also take the form of $\sum\tilde{n}_i$ plus something not independent of this summation, which is quite non-trivial and makes the proof worrying. 
Albeit, in the following, we show that this problem is naturally resolved in our previous discussion. 

Let us focus on $(x-y)+(\lambda(x)-\lambda(y))q$, other factors can be dealt similarly. We can easily see from the defintion that $X$ is connected. 
Therefore we can find some appropriate integers $l$, $i$ and $j$ to relate $x$ and $y$ as $y=x-lp-(j-i)(q-p)-q\sum_{k=i}^{j-1}\tilde{n}'_k$. When we assume that 
the factor under consideration vanishes, we have for some integer $\alpha$, 
\ba
l-(j-i)&=&\alpha q\,,\nn\\
(j-i)+\sum_{k=i}^{j-1}\tilde{n}_k'&=&-(\lambda(x)-\lambda(y))-\alpha p\,.\nn
\ea
Notice that $\sum_{i=1}^N\tilde{n}_i'=p-N$, we can use another integer $j'=j+\alpha$ to rewrite the second equation as
\ba
-(\lambda(x)-\lambda(y))=(j'-i)+\sum_{k=i}^{j'-1}\tilde{n}_k'\,.\nn
\ea
Substituting this back to the relation between $x$ and $y$, we obtain
\ba
y=x-(q-p)(j'-i)-q\sum_{k=i}^{j'-1}\tn'_k-p(j'-i)\,,\nn
\ea
which suggests that if $x\neq y$, we can only have either $x<_X y$ or $x>_X y$. 
The last two cases can again be denied using the definition of the ordering in $X$ as we have done before because additional terms $\sum\tilde{n}'$ cancel out. 

Therefore, the whole discussion in section \ref{s:6.2} applies again, and we see that the $N$-Burge condition provides a generalized prescription from that we proposed in section \ref{section6}. 
Those two conditions (\ref{two-conditions}) here can be treated as a generalized poset construction with which we can count the number of states at each level. 

\section{Conclusion and discussion}
In this paper, we studied some properties of the algebra SH$^c$.
It is supposed to be a universal symmetry which contains
the representations of $\cW_N$-algebra for any $N$.
We show that the singular vectors of a $\cW_N$ module
can be easily understood from the SH$^c$ action on the AFLT basis.
Similar computation leads to the $N$-Burge condition which
characterizes the Hilbert space of the minimal models.

The main focus of the paper is to show
the triality automorphism in the SH$^c$ algebra
which is generated by two mutually noncommuting generators $\sigma_1,\sigma_2$.
While it is the exact automorphism of algebra, it is not straightforward
to see the minimal model CFTs are dual to each other since their rank
are in general different in the second transformation $\sigma_2$.
%While it is invariant under $\sigma_1$, the second generator $\sigma_2$ does not keep
%the form of the central charges except for the cases which correspond to
We have shown an explicit realization of the duality through
the minimal models with the level-rank duality.
Through a diagrammatic representation of primary fields,
we have shown that the infinite number of the central charges of the algebra are kept
invariant after nontrivial cancellation of factors.
It establishes that the minimal models are connected via the triality of SH$^c$.

We also examined the structure of the Hilbert space and found that the duality
is realized as an explicit automorphism between its $N$-tuple Young diagram 
representation and its $M$-tuple dual. 
This morphism is given by shuffling Young diagrams row by row (Figure \ref{linecor}) 
and the basis of the module is the set of all the integer partition $\lambda$ satisfying 
$\lambda(x-N), \lambda(x-M)\le \lambda(x)$, (\ref{Ppartition}).
It is a generalization of the standard partition described by a Young diagram.

Connecting our analysis to the character formulae of minimal models, we have seen that the Hilbert series of P-partitions over a type of partially ordered sets matches the character (\ref{ZvsChi-f}). 
We can prove it explicitly for the Lee-Yang case via the Rogers-Ramanujan identities. 
Conversely, the equation (\ref{ZvsChi-f}) means that, for the Lee-Yang case, the two primary fields give a combinational interpretation of the two Rogers-Ramanujan identities, which was already known in \cite{gordon1961combinatorial}.
Therefore, this equation may give a general combinational interpretation of such identities.
Deeper analysis of the corresponding Hasse diagrams is needed to give a complete proof of (\ref{ZvsChi-f}).

There are some applications of our results to physics.
The immediate one is to $4$D/$2$D correspondence for $\mathcal{N}$=2 supersymmetric gauge theories in 4 dimensions.
On the four-dimensional gauge theory side, the factor $\Lambda^{\pm}$ in (\ref{Lambda}) is 
nothing but the ratio of weights of two fixed point in the instanton moduli space.
Note that we have obtained the resultant representation by dropping some arrangements of 
Young diagrams since we have set the Coulomb branch parameter $(a_{q})_{q=1}^{N}$ and 
the $\Omega$-background parameter $\beta=-\frac{\varepsilon_{2}}{\varepsilon_{1}} $ to specific values. 
This dropping implies that some weights with fixed points vanishes and then, at first sight, 
such fixed points give divergent contributions to the Nekrasov partition function. 
However, this divergence comes from the fact that the fixed points with the torus action are not isolated 
in the moduli space for such a choice of parameters, and then we should not sum all the contributions point by point.
Instead, we should integrate over the set of all the fixed points with an appropriate measure.
It may be interesting to study the correspondence between the sets of all the fixed points 
via the level-rank duality, which might imply a duality between
instanton moduli spaces with two different gauge groups $\mathrm{SU}(N)$ and $\mathrm{SU}(M)$
in the special $\Omega$ background. In a sense, nonisolated fixed points in the equivariant localization will imply the null states of corresponding minimal models.
It may be also interesting to pursue physical meanings of the null states on the $4$D side.

The other application is to higher spin gravity theory, where the universal symmetry
$\cW_\infty[\mu]$ was found.  Since both symmetries describe the minimal models
of $\cW_N$-algebra, SH$^c$ may have some role there too. In particular, it may be interesting
if one can find the role of AFLT basis in the gravitational background.
The other possible application is to the fractional quantum Hall effect (FQH).
Inspired by a mathematical result \cite{feigin2003symmetric},
it was suggested \cite{bernevig2008model} the Jack polynomial may be interpreted as the ground state
wave function. Jack polynomial describes the excited states of the Calogero-Sutherland
model and is identified with AFLT basis with $N=1$ after rewriting the variables by
free boson. If Jack polynomial plays some role, so should be general AFLT basis.
$N$-Burge condition may be interpreted as representing a generalized statistics
for anyon system.

\section*{Acknowledgement}
We are grateful to Yuji Tachikawa for helpful discussions and comments. 
RZ would like to sincerely thank HIROSE international scholarship for generous financial support and K.Goto 
for inspiring consultation. 
YM would like to thank Andrei Mironov  and Alexei Morozov for inviting him
to workshop ``Quantum Geometry, Duality and Matrix Models",
at Lebedev Physical Institute and IITP in Moscow (August 24-30, 2015) where
part of the content of  the paper was presented.
Support from JSPS/RFBR bilateral collaboration ``Faces of matrix models 
in quantum field theory and statistical mechanics" which makes this visit possible
is also gratefully appreciated.
He is obliged to  Tomas Prochazka for the correspondence on the
triality of SH$^c$, Mikhail Bernshtein to point out the relevance of \cite{Feigin:2010qea} and
Hiroaki Kanno for the comments on the poset.
He is partially supported by Grants-in-Aid for Scientific Research (Kakenhi \#25400246).

\appendix
\section{Concrete correspondence of states for $(N,M)=(2,3)$}\label{s:A}

This appendix gives the detailed correspondence between states labeled by various Young diagrams. 
As null states have no correspondence in the dual picture, they certainly will be omitted from the following list. 

For $\Delta=0$ case, the set which labels the rows is $X=\{0,-2,-3,-4,\cdots\}$.
In the following we write $(\lambda(0), \lambda(-2), \lambda(-3),\lambda(-4),\cdots)$
to represent the partition.
At level one (one state):
\ba
\ket{\ydiagram{1},\emptyset}\leftrightarrow\ket{\ydiagram{1},\emptyset,\emptyset}
\leftrightarrow (1,0,0,0,\cdots)\,.\nn
\ea

At level two (three states), 
\ba
&&\ket{\ydiagram{1},\ydiagram{1}}\leftrightarrow\ket{\ydiagram{1,1},\emptyset,\emptyset}
\leftrightarrow (1,0,1,0,\cdots)\,,\quad\nn\\
&&\ket{\ydiagram{1,1},\emptyset}\leftrightarrow\ket{\ydiagram{1},\ydiagram{1},\emptyset}
\leftrightarrow (1,1,0,0,\cdots)\,,\nn\quad\\
&&\ket{\ydiagram{2},\emptyset}\leftrightarrow\ket{\ydiagram{2},\emptyset,\emptyset}
\leftrightarrow (2,0,0,0,\cdots)\,.\nn
\ea

At level three (5 states):
\ba
&&\ket{\ydiagram{1,1},\ydiagram{1}}\leftrightarrow\ket{\ydiagram{1,1},\ydiagram{1},\emptyset}
\leftrightarrow (1,1,1,0,\cdots)\,,\quad \nn\\
&&\ket{\ydiagram{2},\ydiagram{1}}\leftrightarrow\ket{\ydiagram{2,1},\emptyset,\emptyset}
\leftrightarrow (2,0,1,0,\cdots)\,,\quad\nn\\
&& \ket{\ydiagram{1,1,1},\emptyset}\leftrightarrow\ket{\ydiagram{1},\ydiagram{1},\ydiagram{1}}
\leftrightarrow (1,1,0,1,0,\cdots)\,,\nn\\
&&\ket{\ydiagram{2,1},\emptyset}\leftrightarrow\ket{\ydiagram{2},\ydiagram{1},\emptyset}
\leftrightarrow (2,1,0,0,\cdots)\,,\quad\nn\\
&&\ket{\ydiagram{3},\emptyset}\leftrightarrow\ket{\ydiagram{3},\emptyset,\emptyset}
\leftrightarrow (3,0,0,0,\cdots)\,.\nn
\ea

At level four (9 states): (we omit the partition in the following)
\ba
\ket{\ydiagram{1,1,1},\ydiagram{1}}\leftrightarrow\ket{\ydiagram{1,1},\ydiagram{1},\ydiagram{1}}\,,\quad
\ket{\ydiagram{2,1},\ydiagram{1}}\leftrightarrow\ket{\ydiagram{2,1},\ydiagram{1},\emptyset}\,,\quad
\ket{\ydiagram{1,1},\ydiagram{1,1}}\leftrightarrow\ket{\ydiagram{1,1},\ydiagram{1,1},\emptyset}\,,\nn\\
\ket{\ydiagram{3},\ydiagram{1}}\leftrightarrow\ket{\ydiagram{3,1},\emptyset,\emptyset}\,,\quad
\ket{\ydiagram{2},\ydiagram{2}}\leftrightarrow\ket{\ydiagram{2,2},\emptyset,\emptyset}\,,\quad
\ket{\ydiagram{2,1,1},\emptyset}\leftrightarrow\ket{\ydiagram{2},\ydiagram{1},\ydiagram{1}}\,,\nn\\
\ket{\ydiagram{2,2},\emptyset}\leftrightarrow\ket{\ydiagram{2},\ydiagram{2},\emptyset}\,,\quad
\ket{\ydiagram{3,1},\emptyset}\leftrightarrow\ket{\ydiagram{3},\ydiagram{1},\emptyset}\,,\quad
\ket{\ydiagram{4},\emptyset}\leftrightarrow\ket{\ydiagram{4},\emptyset,\emptyset}\,.\nn
\ea

At level five (14 states):
\ba
\ket{\ydiagram{1,1,1,1},\ydiagram{1}}\leftrightarrow\ket{\ydiagram{1,1,1},\ydiagram{1},\ydiagram{1}}\,,\quad
\ket{\ydiagram{2,1,1},\ydiagram{1}}\leftrightarrow\ket{\ydiagram{2,1},\ydiagram{1},\ydiagram{1}}\,,\quad
\ket{\ydiagram{1,1,1},\ydiagram{1,1}}\leftrightarrow\ket{\ydiagram{1,1},\ydiagram{1,1},\ydiagram{1}}\,,\nn\\
\ket{\ydiagram{2,2},\ydiagram{1}}\leftrightarrow\ket{\ydiagram{2,1},\ydiagram{2},\emptyset}\,,\quad
\ket{\ydiagram{3,1},\ydiagram{1}}\leftrightarrow\ket{\ydiagram{3,1},\ydiagram{1},\emptyset}\,,\quad
\ket{\ydiagram{2,1},\ydiagram{1,1}}\leftrightarrow\ket{\ydiagram{2,1},\ydiagram{1,1},\emptyset}\,,\nn\\
\ket{\ydiagram{2,1},\ydiagram{2}}\leftrightarrow\ket{\ydiagram{2,2},\ydiagram{1},\emptyset}\,,\quad
\ket{\ydiagram{4},\ydiagram{1}}\leftrightarrow\ket{\ydiagram{4,1},\emptyset,\emptyset}\,,\quad
\ket{\ydiagram{3},\ydiagram{2}}\leftrightarrow\ket{\ydiagram{3,2},\emptyset,\emptyset}\,,\nn\\
\ket{\ydiagram{2,2,1},\emptyset}\leftrightarrow\ket{\ydiagram{2},\ydiagram{2},\ydiagram{1}}\,,\quad
\ket{\ydiagram{3,1,1},\emptyset}\leftrightarrow\ket{\ydiagram{3},\ydiagram{1},\ydiagram{1}}\,,\quad
\ket{\ydiagram{3,2},\emptyset}\leftrightarrow\ket{\ydiagram{3},\ydiagram{2},\emptyset}\nn\\
\ket{\ydiagram{4,1},\emptyset}\leftrightarrow\ket{\ydiagram{4},\ydiagram{1},\emptyset}\,,\quad
\ket{\ydiagram{5},\emptyset}\leftrightarrow\ket{\ydiagram{5},\emptyset,\emptyset}\,.\nn
\ea

For $\Delta=-1/5$, $X=\{-1,-2,-3,-4,\dots\}$. We have two states at the first level:
\ba
\ket{\ydiagram{1},\emptyset}\leftrightarrow\ket{\emptyset,\ydiagram{1},\emptyset}\leftrightarrow(1,0,0,0,\dots)\,,\nn\\
\ket{\emptyset,\ydiagram{1}}\leftrightarrow\ket{\ydiagram{1},\emptyset,\emptyset}\leftrightarrow(0,1,0,0,\dots)\,.\nn
\ea

At level two (four states): (we omit the partition in the following)
\ba
\ket{\ydiagram{1,1},\emptyset}\leftrightarrow\ket{\emptyset,\ydiagram{1},\ydiagram{1}}\leftrightarrow(1,0,1,0,\dots)\,,\nn\\
\ket{\ydiagram{2},\emptyset}\leftrightarrow\ket{\emptyset,\ydiagram{2},\emptyset}\leftarrow(2,0,0,0,\dots)\,,\nn\\
\ket{\ydiagram{1},\ydiagram{1}}\leftrightarrow\ket{\ydiagram{1},\ydiagram{1},\emptyset}\leftrightarrow(1,1,0,0,\dots)\,,\nn\\
\ket{\emptyset,\ydiagram{2}}\leftrightarrow\ket{\ydiagram{2},\emptyset,\emptyset}\leftrightarrow(0,2,0,0,\dots)\,.\nn
\ea

At level three (seven states):
\ba
\ket{\ydiagram{1,1},\ydiagram{1}}\leftrightarrow\ket{\ydiagram{1},\ydiagram{1},\ydiagram{1}}\leftrightarrow(1,1,1,0,\dots)\,,\nn\\
\ket{\ydiagram{2,1},\emptyset}\leftrightarrow\ket{\emptyset,\ydiagram{2},\ydiagram{1}}\leftrightarrow(2,0,1,0,\dots)\,,\nn\\
\ket{\ydiagram{3},\emptyset}\leftrightarrow\ket{\emptyset,\ydiagram{3},\emptyset}\leftrightarrow(3,0,0,0,\dots)\,,\nn\\
\ket{\ydiagram{2},\ydiagram{1}}\leftrightarrow\ket{\ydiagram{1},\ydiagram{2},\emptyset}\leftrightarrow(2,1,0,0,\dots)\,,\nn\\
\ket{\ydiagram{1},\ydiagram{1,1}}\leftrightarrow\ket{\ydiagram{1},\ydiagram{1,1},\emptyset}\leftrightarrow(1,1,0,1,\dots)\,,\nn\\
\ket{\ydiagram{1},\ydiagram{2}}\leftrightarrow\ket{\ydiagram{2},\ydiagram{1},\emptyset}\leftrightarrow(1,2,0,0,\dots)\,,\nn\\
\ket{\emptyset,\ydiagram{3}}\leftrightarrow\ket{\ydiagram{3},\emptyset,\emptyset}\leftrightarrow(0,3,0,0,\dots)\,.\nn
\ea

At level four (13 states):
\ba
\ket{\ydiagram{4},\emptyset}\leftrightarrow\ket{\emptyset,\ydiagram{4},\emptyset}\,,\quad
\ket{\ydiagram{3,1},\emptyset}\leftrightarrow\ket{\emptyset,\ydiagram{3},\ydiagram{1}}\,,\quad
\ket{\ydiagram{3},\ydiagram{1}}\leftrightarrow\ket{\ydiagram{1},\ydiagram{3},\emptyset}\,,\nn\\
\ket{\ydiagram{2,1},\ydiagram{1}}\leftrightarrow\ket{\ydiagram{1},\ydiagram{2},\ydiagram{1}}\,,\quad
\ket{\ydiagram{2,2},\emptyset}\leftrightarrow\ket{\emptyset,\ydiagram{2},\ydiagram{2}}\,,\quad
\ket{\ydiagram{2},\ydiagram{1,1}}\leftrightarrow\ket{\ydiagram{1},\ydiagram{2,1},\emptyset}\nn\\
\ket{\ydiagram{2},\ydiagram{2}}\leftrightarrow\ket{\ydiagram{2},\ydiagram{2},\emptyset}\,,\quad
%\ket{\ydiagram{2},\ydiagram{1,1}}\leftrightarrow\ket{\ydiagram{1},\ydiagram{2,1},\emptyset}\nn\\                                                                                                                                                                                        
\ket{\ydiagram{1,1},\ydiagram{1,1}}\leftrightarrow\ket{\ydiagram{1},\ydiagram{1,1},\ydiagram{1}}\,,\quad
\ket{\ydiagram{1,1,1},\ydiagram{1}}\leftrightarrow\ket{\ydiagram{1,1},\ydiagram{1},\ydiagram{1}}\,,\nn\\
\ket{\ydiagram{1,1},\ydiagram{2}}\leftrightarrow\ket{\ydiagram{2},\ydiagram{1},\ydiagram{1}}\,,\quad
\ket{\ydiagram{1},\ydiagram{2,1}}\leftrightarrow\ket{\ydiagram{2},\ydiagram{1,1},\emptyset}\,,\quad
\ket{\ydiagram{1},\ydiagram{3}}\leftrightarrow\ket{\ydiagram{3},\ydiagram{1},\emptyset}\,,\nn\\
\ket{\emptyset,\ydiagram{4}}\leftrightarrow\ket{\ydiagram{4},\emptyset,\emptyset}\,.\nn
\ea

At level five: (21 states)
\ba
\ket{\ydiagram{5},\emptyset}\leftrightarrow\ket{\emptyset,\ydiagram{5},\emptyset}\,,\quad
\ket{\ydiagram{4,1},\emptyset}\leftrightarrow\ket{\emptyset,\ydiagram{4},\ydiagram{1}}\,,\quad
\ket{\ydiagram{4},\ydiagram{1}}\leftrightarrow\ket{\ydiagram{1},\ydiagram{4},\emptyset}\,,\nn\\
\ket{\ydiagram{3,1},\ydiagram{1}}\leftrightarrow\ket{\ydiagram{1},\ydiagram{3},\ydiagram{1}}\,,\quad
\ket{\ydiagram{3},\ydiagram{1,1}}\leftrightarrow\ket{\ydiagram{1},\ydiagram{3,1},\emptyset}\,,\quad
\ket{\ydiagram{3},\ydiagram{2}}\leftrightarrow\ket{\ydiagram{2},\ydiagram{3},\emptyset}\,,\nn\\
\ket{\ydiagram{3,2},\emptyset}\leftrightarrow\ket{\emptyset,\ydiagram{3},\ydiagram{2}}\,,\quad
\ket{\ydiagram{2,2},\ydiagram{1}}\leftrightarrow\ket{\ydiagram{1},\ydiagram{2},\ydiagram{2}}\,,\quad
\ket{\ydiagram{2,1,1},\ydiagram{1}}\leftrightarrow\ket{\ydiagram{1,1},\ydiagram{2},\ydiagram{1}}\,,\nn\\
\ket{\ydiagram{2},\ydiagram{3}}\leftrightarrow\ket{\ydiagram{3},\ydiagram{2},\emptyset}\,,\quad
\ket{\ydiagram{2,1},\ydiagram{1,1}}\leftrightarrow\ket{\ydiagram{1},\ydiagram{2,1},\ydiagram{1}}\,,\quad
\ket{\ydiagram{2,1},\ydiagram{2}}\leftrightarrow\ket{\ydiagram{2},\ydiagram{2},\ydiagram{1}}\,,\nn\\
\ket{\ydiagram{2},\ydiagram{2,1}}\leftrightarrow\ket{\ydiagram{2},\ydiagram{2,1},\emptyset}\,,\quad
\ket{\ydiagram{1,1,1},\ydiagram{2}}\leftrightarrow\ket{\ydiagram{2,1},\ydiagram{1},\ydiagram{1}}\,,\quad
\ket{\ydiagram{1,1,1},\ydiagram{1,1}}\leftrightarrow\ket{\ydiagram{1,1},\ydiagram{1,1},\ydiagram{1}}\,,\nn\\
\ket{\ydiagram{1,1},\ydiagram{1,1,1}}\leftrightarrow\ket{\ydiagram{1},\ydiagram{1,1},\ydiagram{1,1}}\,,\quad
\ket{\ydiagram{1,1},\ydiagram{2,1}}\leftrightarrow\ket{\ydiagram{2},\ydiagram{1,1},\ydiagram{1}}\,,\quad
\ket{\ydiagram{1,1},\ydiagram{3}}\leftrightarrow\ket{\ydiagram{3},\ydiagram{1},\ydiagram{1}}\,,\nn\\
\ket{\ydiagram{1},\ydiagram{3,1}}\leftrightarrow\ket{\ydiagram{3},\ydiagram{1,1},\emptyset}\,,\quad
\ket{\ydiagram{1},\ydiagram{4}}\leftrightarrow\ket{\ydiagram{4},\ydiagram{1},\emptyset}\,,\quad
\ket{\emptyset,\ydiagram{5}}\leftrightarrow\ket{\ydiagram{5},\emptyset,\emptyset}\,.\nn
\ea

\bibliographystyle{utphys}
\bibliography{FNMZ}

\end{document}